\documentclass[preprint]{aastex62}

\usepackage{bm}
\usepackage{threeparttable}
\usepackage{amsmath,amssymb}
\usepackage{textcomp}
\usepackage{booktabs}
\usepackage{natbib}
\usepackage{chapterbib}
\usepackage{comment}
\usepackage{here}
\usepackage {threeparttable}

\shortauthors{Kondo et al.}

\begin{document}

\title{MOA-bin-29b : A Microlensing Gas-giant Planet Orbiting a Low-mass Host Star}

\author{Iona Kondo}
\affiliation{MOA collaboration}
\affiliation{Department of Earth and Space Science, Graduate School of Science, Osaka University, Toyonaka, Osaka 560-0043, Japan}

\author{Takahiro Sumi}
\affiliation{MOA collaboration}
\affiliation{Department of Earth and Space Science, Graduate School of Science, Osaka University, Toyonaka, Osaka 560-0043, Japan}

\author{David P.~Bennett}
\affiliation{MOA collaboration}
\affiliation{Code 667, NASA Goddard Space Flight Center, Greenbelt, MD 20771, USA}
\affiliation{Department of Astronomy, University of Maryland, College Park, MD 20742, USA}

\author{Andrzej Udalski}
\affiliation{OGLE collaboration}
\affiliation{Warsaw University Observatory, Al. Ujazdowskie 4, 00-478 Warszawa, Poland}

\author{Ian A. Bond}
\affiliation{MOA collaboration}
\affiliation{Institute of Natural and Mathematical Sciences, Massey University, Auckland 0745, New Zealand}

\author{Nicholas J. Rattenbury}
\affiliation{MOA collaboration}
\affiliation{Department of Physics, University of Auckland, Private Bag 92019, Auckland, New Zealand}

\author{Valerio Bozza}
\affiliation{MiNDSTEp Collaboration}
\affiliation{Dipartimento di Fisica “E.R. Caianiello”, Universit\`{a} di Salerno, Via Giovanni Paolo II 132, 84084, Fisciano, Italy}
\affiliation{Istituto Nazionale di Fisica Nucleare, Sezione di Napoli, Napoli, Italy}

\author{Yuki Hirao}
\affiliation{MOA collaboration}
\affiliation{Department of Earth and Space Science, Graduate School of Science, Osaka University, Toyonaka, Osaka 560-0043, Japan}

\author{Daisuke Suzuki}
\affiliation{MOA collaboration}
\affiliation{Institute of Space and Astronautical Science, Japan Aerospace Exploration Agency, 3-1-1 Yoshinodai, Chuo, Sagamihara, Kanagawa, 252-5210, Japan}

\author{Naoki Koshimoto}
\affiliation{MOA collaboration}
\affiliation{Department of Astronomy, Graduate School of Science, The University of Tokyo, 7-3-1 Hongo, Bunkyo-ku, Tokyo 113-0033, Japan}
\affiliation{National Astronomical Observatory of Japan, 2-21-1 Osawa, Mitaka, Tokyo 181-8588, Japan}

\author{Masayuki Nagakane}
\affiliation{MOA collaboration}
\affiliation{Department of Earth and Space Science, Graduate School of Science, Osaka University, Toyonaka, Osaka 560-0043, Japan}

\author{Shota Miyazaki}
\affiliation{MOA collaboration}
\affiliation{Department of Earth and Space Science, Graduate School of Science, Osaka University, Toyonaka, Osaka 560-0043, Japan}
\nocollaboration

\author{F. Abe}
\affiliation{Institute for Space-Earth Environmental Research, Nagoya University, Nagoya 464-8601, Japan}
\author{R. Barry}
\affiliation{Code 667, NASA Goddard Space Flight Center, Greenbelt, MD 20771, USA}
\author{A. Bhattacharya}
\affiliation{Code 667, NASA Goddard Space Flight Center, Greenbelt, MD 20771, USA}
\affiliation{Department of Astronomy, University of Maryland, College Park, MD 20742, USA}
\author{M. Donachie}
\affiliation{Department of Physics, University of Auckland, Private Bag 92019, Auckland, New Zealand}
\author{A. Fukui}
\affiliation{Department of Earth and Planetary Science, Graduate School of Science, The University of Tokyo, 7-3-1 Hongo, Bunkyo-ku, Tokyo 113-0033, Japan}
\affiliation{Instituto de Astrof\'isica de Canarias, V\'ia L\'actea s/n, E-38205 La Laguna, Tenerife, Spain}
\author{H. Fujii}
\affiliation{Institute for Space-Earth Environmental Research, Nagoya University, Nagoya 464-8601, Japan}
\author{Y. Itow}
\affiliation{Institute for Space-Earth Environmental Research, Nagoya University, Nagoya 464-8601, Japan}
\author{Y. Kamei}
\affiliation{Institute for Space-Earth Environmental Research, Nagoya University, Nagoya 464-8601, Japan}
\author{M. C. A. Li}
\affiliation{Department of Physics, University of Auckland, Private Bag 92019, Auckland, New Zealand}
\author{Y. Matsubara}
\affiliation{Institute for Space-Earth Environmental Research, Nagoya University, Nagoya 464-8601, Japan}
\author{T. Matsuo}
\affiliation{Department of Earth and Space Science, Graduate School of Science, Osaka University, Toyonaka, Osaka 560-0043, Japan}
\author{Y. Muraki}
\affiliation{Institute for Space-Earth Environmental Research, Nagoya University, Nagoya 464-8601, Japan}
\author{C. Ranc}
\affiliation{Code 667, NASA Goddard Space Flight Center, Greenbelt, MD 20771, USA}
\author{H. Shibai}
\affiliation{Department of Earth and Space Science, Graduate School of Science, Osaka University, Toyonaka, Osaka 560-0043, Japan}
\author{H. Suematsu}
\affiliation{Department of Earth and Space Science, Graduate School of Science, Osaka University, Toyonaka, Osaka 560-0043, Japan}
\author{D. J. Sullivan}
\affiliation{School of Chemical and Physical Sciences, Victoria University, Wellington, New Zealand}
\author{P. J. Tristram}
\affiliation{University of Canterbury Mt.\ John Observatory, P.O. Box 56, Lake Tekapo 8770, New Zealand}
\author{T. Yamakawa}
\affiliation{Institute for Space-Earth Environmental Research, Nagoya University, Nagoya 464-8601, Japan}
\author{A. Yonehara}
\affiliation{Department of Physics, Faculty of Science, Kyoto Sangyo University, 603-8555 Kyoto, Japan}
\collaboration{(MOA collaboration)}

\author{P. Mr{\'o}z}
\affiliation{Warsaw University Observatory, Al. Ujazdowskie 4, 00-478 Warszawa, Poland}
\author{M.~K. Szyma{\'n}ski}
\affiliation{Warsaw University Observatory, Al. Ujazdowskie 4, 00-478 Warszawa, Poland}
\author{I. Soszy{\'n}ski}
\affiliation{Warsaw University Observatory, Al. Ujazdowskie 4, 00-478 Warszawa, Poland}
\author{K. Ulaczyk}
\affiliation{Warsaw University Observatory, Al. Ujazdowskie 4, 00-478 Warszawa, Poland}
\affiliation{Department of Physics, University of Warwick, Gibbet Hill Road, Coventry, CV4 7AL, UK}
\collaboration{(OGLE collaboration)}

\begin{abstract}
We report the discovery of a gas-giant planet orbiting a low-mass host star in the microlensing event MOA-bin-29 that occurred in 2006. 
We find five degenerate solutions with the planet/host-star mass ratio of $q \sim 10^{-2}$. The Einstein radius crossing time of all models are relatively short ($\sim 4-7$ days), which indicates that the mass of host star is likely low.
The measured lens-source proper motion is $5-9$ ${\rm mas}\ {\rm yr}^{-1}$ depending on the models.
Since only finite source effects are detected, we conduct a Bayesian analysis in order to obtain the posterior probability distribution of the lens physical properties. 
As a result, we find the lens system is likely to be a gas-giant orbiting a brown dwarf or a very late M-dwarf in the Galactic bulge.
The probability distributions of the physical parameters for the five degenerate models are consistent within the range of error.
By combining these probability distributions, we conclude that the lens system is a gas giant with a mass of $M_{\rm p} = 0.63^{+1.13}_{-0.39}\ M_{\rm Jup}$ orbiting a brown dwarf with a mass of $M_{\rm h} = 0.06^{+0.11}_{-0.04}\ M_\odot$ at a projected star--planet separation of $r_\perp = 0.53^{+0.89}_{-0.18}\ {\rm au}$. The lens distance is $D_{\rm L} = 6.89^{+1.19}_{-1.19}\ {\rm kpc}$, i.e., likely within the Galactic bulge.
\end{abstract}
\keywords{gravitational lensing: micro --- planets and satellites: detection}

\section{Introduction}
\label{sec-intr}
More than 3900 exoplanets have been discovered since the first discovery of an exoplanet orbiting a main-sequence star in 1995 \citep{1995Natur.378..355M}, including various planetary systems, such as hot Jupiters and super Jupiters.  
Most known exoplanets have been found by the radial velocity \citep{2006ApJ...646..505B, 2011A&A...534A..58P} and transit methods \citep{2013ApJS..204...24B}, which are most sensitive to massive planets in close orbits. Direct imaging has found young giant planets in very wide orbits.
\\

The gravitational microlensing method has a unique planet sensitivity to planets down to low masses \citep{1996ApJ...472..660B} in wide orbits, just beyond the snow line \citep{1992ApJ...396..104G}. Exoplanet searches by using the microlensing were first proposed by \citet{1991ApJ...374L..37M}, and over 90 planets have been discovered by this method to date.
Gravitational microlensing occurs when a foreground lens star crosses the line of sight between an observer and a background source star by chance. 
The gravity of the lens star bends the light from the source star and magnifies its brightness. If the lens star has a companion, its gravity affects the magnification of the source star.
The microlensing method does not depend on the brightness of the lens objects. So we can discover low-mass companions around faint and/or distant host, such as M-dwarfs or even brown dwarfs in the Galactic disk and bulge. 
\\

The formation theory of gas giants around the low-mass host remains to be fully elucidated. 
According to the core accretion theory, it is difficult to form gas-giant planets in the disks around low-mass stars \citep{1996Icar..124...62P, 2004ApJ...616..567I, 2004ApJ...612L..73L, 2006ApJ...650L.139K}.  
Gravitational instability in the protoplanetary disk may play the important role in the formation of gas giants \citep{2006ApJ...643..501B}. In order to constrain the formation theory, more observational samples with low-mass hosts are required. 
By using the microlensing method, many planetary systems with low-mass stars have been discovered in orbital separation between $\sim0.2-10$ au \citep{2013ApJ...763...67S, 2015ApJ...804...33S, 2017AJ....154...35N, 2018AJ....155..219J}. This is complementary to the other detection methods.
\\

In this paper, we present the analysis of the planetary microlensing event MOA-bin-29 with a short Einstein radius crossing time of $t_{\rm E} \sim 4-7$ days, which suggests the host is a low-mass object. Section \ref{sec-obs} explains observations and data. 
Our light-curve modeling method and result are shown in Section \ref{sec-model}. 
In Section \ref{sec-cmd}, we derive an angular Einstein radius from the source magnitude and color.
In Section \ref{sec-bay}, physical parameters of the lens system are estimated with a Bayesian analysis.
Finally, we discuss our analysis and reach conclusions in Section \ref{sec-dis}.
\\

\section{Observations}
\label{sec-obs}
The Microlensing Observations in Astrophysics (MOA; \citealp{2001MNRAS.327..868B,2003ApJ...591..204S}) collaboration conducts a microlensing exoplanet survey toward the Galactic bulge by using the 1.8m MOA-II telescope with a 2.2 ${\rm deg^{2}}$ wide field-of-view (FOV) CCD-camera, MOA-cam3 \citep{2008ExA....22...51S} at Mt. John University Observatory in New Zealand. Thanks to the wide FOV, a high-cadence survey observation can be conducted. 
MOA survey uses a custom wide-band filter referred as $R_{\rm MOA}$, corresponding to a Cousins $R$- and $I$-band.
The MOA photometry is reduced by using the MOA's implementation of the Difference Image Analysis (DIA) pipeline \citep{2001MNRAS.327..868B}. 
\\ 

The Optical Gravitational Lensing Experiment (OGLE; \citealp{2003AcA....53..291U} ) also conducts a microlensing survey at the Las Campanas Observatory in Chile. The third phase of the survey, OGLE-III used the 1.3m Warsaw telescope with a 0.35 ${\rm deg^{2}}$ FOV CCD-camera. Currently, its forth phase, OGLE-IV \citep{2015AcA....65....1U} started its high-cadence survey in 2010 with a 1.4 ${\rm deg^{2}}$ FOV CCD-camera. The OGLE photometry is reduced by the OGLE's implementation of the DIA photometry pipeline \citep{2003AcA....53..291U}.
\\ 

\begin{figure}[h]
\begin{center}
     \includegraphics[scale=0.5, angle=270]{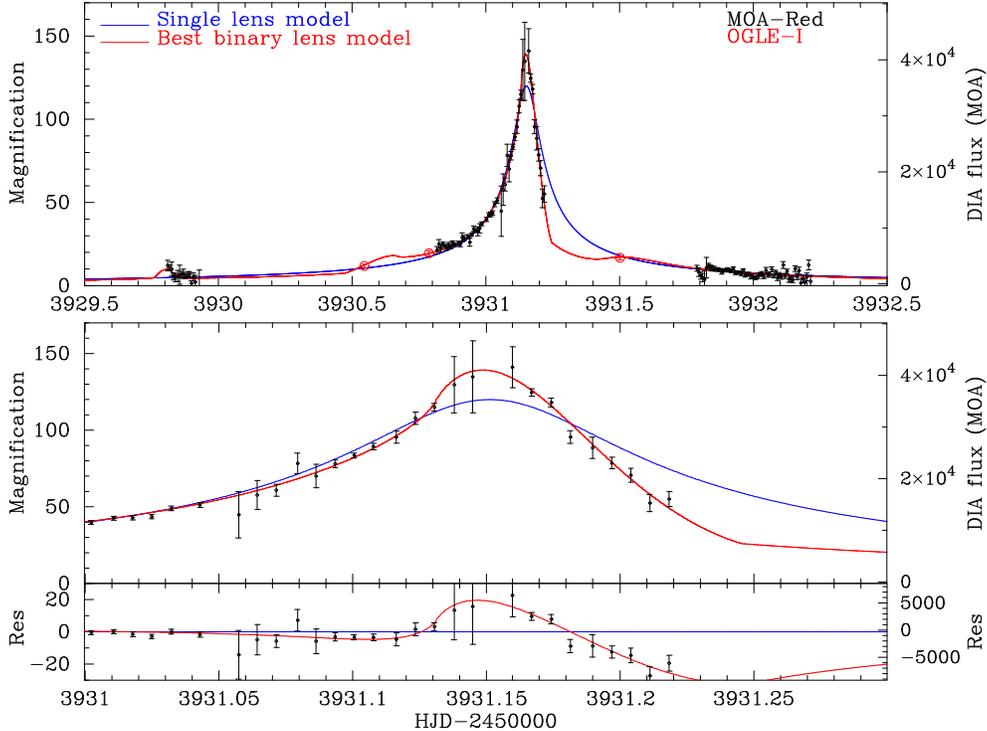}
   \caption{The top panel shows the light curve and models for MOA-bin-29. The vertical axis on the right shows the DIA flux of MOA. The blue line shows the single-lens model, and the red line shows the best planetary model. The middle panel shows the light curve around the peak. We can find clear deviation of data points from the single-lens model. The bottom panel shows the residuals from the single-lens model.}
   \label{fig-pspl}
\end{center}
\end{figure}

\begin{figure}[h]
\begin{center}
   \includegraphics[scale=0.30, angle=90]{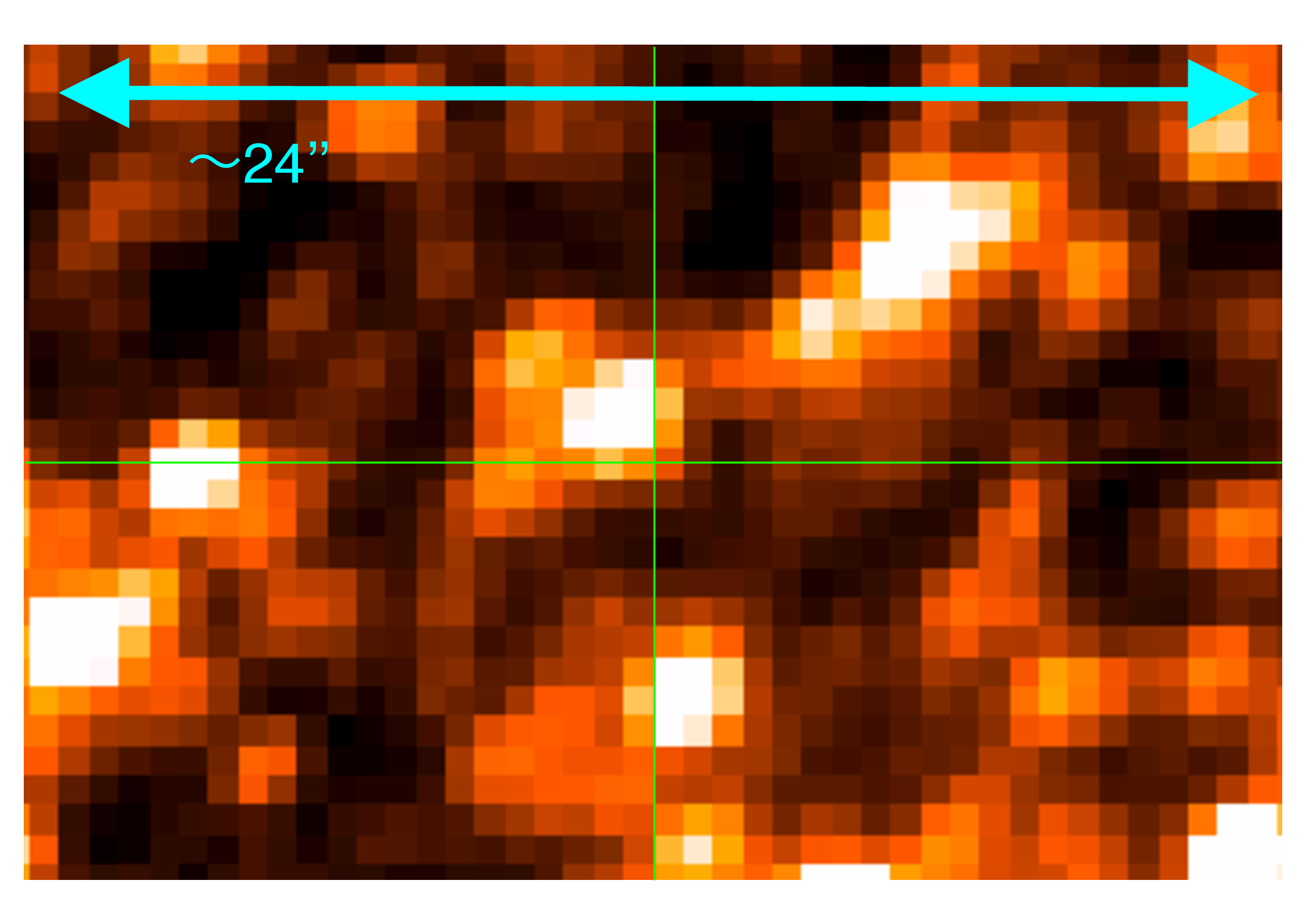}
   \caption{The reference image around MOA-bin-29. The green cross shows the position of the event. North is up and east is to the left.}
   \label{ref}
\end{center}
\end{figure}

The gravitational microlensing event MOA-bin-29 reached to near peak on 2006 July 14 (${\rm HJD'}$ $=$ ${\rm HJD} - 2450000$ $\sim$ $3929.77$), at the J2000 equatorial coordinates $({\rm RA, Dec})$ = ($17^{h}$ $57^{m}$ $30^{s}$.$23, -29^{\circ}$ $44^{\prime}$ $11^{\prime \prime}.63$) corresponding to Galactic coordinates $(l, b)$ = $(0.633^{\circ}, -2.636^{\circ})$. Figure \ref{fig-pspl} shows the MOA and OGLE-III light curves.
This event was not detected in the real-time analysis but was found only after the off-line analysis of MOA database during $2006-2014$ (Sumi et al. 2019, in preparation).
There are several possible reasons why this event was not detected by the MOA Alert system. 
First, this was a short duration event.
Second, the alert system had just started since 2006 and the baseline was not long enough to distinguish from other variables. 
In this off-line analysis, the $2006-2014$ MOA Galactic bulge data have been re-analyzed and the events were detected using a criteria that is different from MOA alert system.
Since this event was not alerted, there were no follow-up observations and only the survey data is available.
Fortunately, however, the event is located in MOA field gb9, which was observed with the highest cadence of 10 minutes, so we have good coverage of this short event.
The event is also located in the OGLE-III field BLG102, and we obtained data covering some part of the light curve during its magnification. 
Consequently, we must characterize the anomaly with only the survey groups' observations.
Figure \ref{ref} shows the reference image around MOA-bin-29. 
The green cross indicates the position of the event detected on the difference images. We found that the source star is much fainter than a nearby bright star (located up and to the left of the event location in Figure \ref{ref}).
\\ 


The photometric error bars produced by the data pipelines can be underestimated (or more rarely overestimated).
We should consider other systematic errors caused by observational conditions and so on. In order to get proper errors of the parameters in the light-curve modeling, we empirically normalize the error bars by using the standard method of \citep{2008ApJ...684..663B}.
We use the formula,
\begin{equation}
  \sigma^{\prime}_{i} = k \sqrt{\sigma^{2}_{i} + e^{2}_{\rm min}},
\end{equation}
where $\sigma^{\prime}_{i}$ is the $i$ th renormalized error, $\sigma_{i}$ is the $i$ th error obtained from DIA, and $k$ and $e_{\rm min}$ are the renormalizing parameters. 
We set $e_{\rm min}=0.003$ to account for flat-fielding errors, and we adjust the value of $k$ and  $\chi^2{\rm /dof}=1$. The normalization parameters and the number of data of each telescope are given in Table  \ref{data}. We use the MOA data for $2006-2014$ and the OGLE data for $2006-2009$.
\\

\begin{table}[htbp]
\caption{The number of data points in the light curves and the normalization parameters}
\label{data}
\begin{center}
 \begin{tabular}{llrrr}
 \hline \hline
Telescope & Filter          & $k$ & $e_{\rm min}$          &  Number of Data                                       \\  \hline 
 MOA  & $R_{\rm MOA}$         & 1.105  & 0.003   &  29094                  \\           
 OGLE   & $I$      &  1.058  & 0.003   &     871   \\  
 \hline \hline
  \end{tabular}
  \end{center}
\end{table}

\begin{figure}[h]
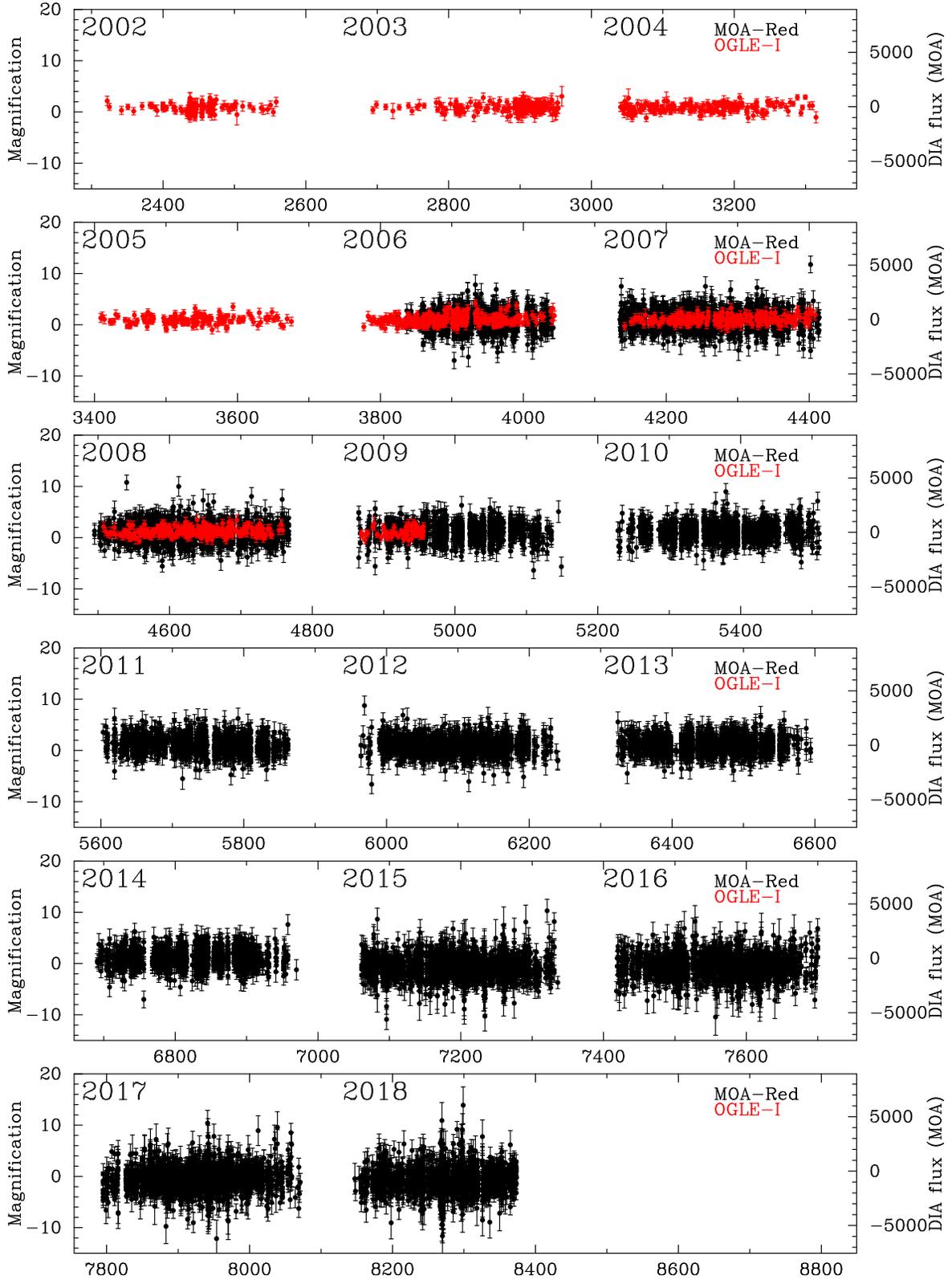

 \begin{minipage}{1\hsize}
 	\centering
	\includegraphics[scale=0.6,angle=270]{baseline0608_2002.ps}
\end{minipage}
\begin{minipage}{1\hsize}
 	\centering
	\includegraphics[scale=0.6,angle=270]{baseline0608_2005.ps}
\end{minipage}
 \begin{minipage}{1\hsize}
 	\centering
       \includegraphics[scale=0.6,angle=270]{baseline0608_2008.ps}
\end{minipage}
\begin{minipage}{1\hsize}
 	\centering
	\includegraphics[scale=0.6,angle=270]{baseline0608_2011.ps}
\end{minipage}
 \begin{minipage}{1\hsize}
 	\centering
	\includegraphics[scale=0.6,angle=270]{baseline0608_2014.ps}
\end{minipage}
 \begin{minipage}{1\hsize}
 	\centering
	\includegraphics[scale=0.6,angle=270]{baseline0608_2017.ps}
\end{minipage}

  \caption{The baseline of the full light curve for $2002-2018$, where the MOA-II data for $2006-2018$ and the OGLE-III data for $2002-2009$. The vertical axis on the right shows the DIA flux of MOA and the left vertical axis shows the magnification corresponding to the Figure \ref{fig-pspl}. The data points during the event on $3929.5 < {\rm HJD'} < 3932.5$ are not shown. The MOA data points with flux errors $> 1000$ ADU or with seeing $> 4$ pixels are omitted for clarity. }
  \label{fig-baseline}
 \end{figure}

We investigated the possibility that this short magnification is not due to the microlensing, but the other artifacts or intrinsic variability of the star \citep{2011Natur.473..349S, 2012ApJ...757..119B}. First, we examined the pixel level DIA images of the target and confirmed that the event is not due to the fast moving objects nor cosmic ray hits. Second, we show the baseline of the full light curve except during the event at $3929.5 < {\rm HJD'} < 3932.5$ in Figure \ref{fig-baseline}. Here, we omitted the MOA data points with flux errors $> 1000$ ADU or with seeing $> 4$ pixels for clarity.  We found that there are no other obvious magnifications in the baseline for $2002-2018$. This indicates that the event does not likely consist of a cataclysmic variables (CVs) nor flare stars because most of them repeat in a timescale of a few years or less. Additionally, the light-curve shape of these flare events is usually fast-rise and slow-decline, while the MOA-bin-29 light curve shows the opposite, i.e., slow-rise and fast-decline. Furthermore, the variables that can be modeled by binary microlensing light curves tend to have physically unlikely parameters and a small number of magnified data points \citep{2012ApJ...757..119B} and neither of these conditions apply to MOA-bin-29. These considerations indicate that this event is a microlensing event.

\section{Light-curve Models}
\label{sec-model}
The single-lens light-curve model depends on three parameters: the time of lens-source closest approach $t_{0}$, the Einstein ring crossing time $t_{\rm E}$, and the impact parameter in units of the Einstein radius $u_{0}$. Binary-lens models require four additional parameters: the planet-host mass ratio, $q$, the planet-host separation in units of the Einstein radius, $s$, the angle between the trajectory of the source and the planet-host axis, $\alpha$, and the ratio of the angular source size to the angular Einstein radius. If we were to include microlensing parallax, we would need two additional parameters.
The model flux $F(t)$ of magnified source as a function of time $t$ can be given by, 
\begin{equation}
  F(t) = A(t)F_S + F_b,
  \label{eq-FsFb}
\end{equation}
where $A(t)$ is a magnification of the source flux at $t$, $F_S$ is the baseline flux of the source star, and $F_b$ is the baseline flux of any unresolved light.
The flux on the DIA image, $\Delta$ $F(t)$, is the difference between $F(t)$ and the flux on the reference image at $t_{\rm ref}$, $F(t_{\rm ref})={\rm const.}$, and is given by
\begin{eqnarray}
  \Delta F(t) &=& F(t) - F(t_{\rm ref}) = A(t)F_S + F_b', \\
  F_b' &=& F_b - F(t_{\rm ref}).
\end{eqnarray}
The large number of the parameters of the microlensing event and correlations with each parameter make 
it difficult to search for the best-fit model parameters. The number of nonlinear parameters can be reduced
by noting that equation~\ref{eq-FsFb} is linear in $F_S$ and $F_B$, so these parameters can be solved for
directly for any $F(t)$ that is considered \citep{rhie_98smc1}. 

The initial modeling of this event that led to its classification as a likely planetary microlensing event was
done using three different modeling methods, using the modeling codes of \citet{2010ApJ...716.1408B},
\citet{bozza10}, and \citet{2010ApJ...710.1641S}. Of particular importance for an event like MOA-bin-29, which 
does not have a significant signal from part of the light curve that resembles a single-lens light curve, is the
variation of the \citet{2010ApJ...716.1408B} method described in \citet{2012ApJ...757..119B}.
This search code variation centers the grid search on the centers of caustics of wide or close binaries
that are widely separated from the central caustic, and is able to efficiently find the correct solutions
for events dominated by non-central caustics, such as MOA-bin-1 (dominated by a wide planetary
caustic) and MOA-bin-3 (dominated by a minor image caustic of a close binary-lens system). This analysis
found that only central caustic models were competitive within $\Delta\chi^2$ of 40.

After this preliminary analysis ruled out the widely separated caustic models, we used the \citet{2010ApJ...710.1641S}
code for our detailed analysis. This code combines the Markov Chain Monte Carlo (MCMC) algorithm 
\citep{2003ApJS..148..195V}  with the image-centered ray-shooting method 
\citep{1996ApJ...472..660B, 2010ApJ...716.1408B}. 
First, we performed a broad grid search over the at 9680 different grid points, over $(q, s, \alpha)$ space 
with other parameters free. 
Next, we refined all parameters for the best 100 models with the smallest $\chi^{2}$ to search for the global best-fit model.
In conducting our grid search, we have set the initial parameters with range of $t_{0} \pm 10$ days, and $u_{0} \pm 0.5$.
This, our grid search covers a wider parameter space. Furthermore, when we refine the model by 
freeing the $q$, $s$, and $\alpha$ parameters, the
MCMC will probe for the wider solutions outside of each local minima. 
For example, the MCMC can find the solution with $q>1$ where $t_{0}$ becomes
the time of the magnification near the secondary companion.

\subsection{Limb Darkening}
Binary-lens events usually have caustic crossings of cusp approaches that resolve the finite angular size of the source, so we must include the limb darkening of the source star.
In order to take these effects into account, we adopt the following linear limb-darkening law:
\begin{equation}
S_\lambda(\vartheta) = S_\lambda(0)[1-u_\lambda(1-\cos (\vartheta))],
\end{equation}
where $S_\lambda(\vartheta)$ is a limb-darkening surface brightness.
The effective temperature of the source star estimated from the extinction-free source color presented in Section \ref{sec-cmd} is $T_{\rm eff} \sim 4939 \, \rm K$ \citep{2009A&A...497..497G}. 
Assuming the surface gravity $\log g = 4.5$ and metallicity of $\log [{\rm M/H}] = 0$, we find limb-darkening coefficients of $u_I = 0.5880$ and $u_R = 0.6809$ from the ATLAS model \citep{2011A&A...529A..75C}. For the $R_{\rm MOA}$ passband, we use the coefficient for $u_{\rm Red} = 0.6345$, which is the mean of $u_I$ and $u_R$.
\\

\subsection{Best-fit model and Degenerate models}
\label{secsub-model}
\begin{figure}[h]
\begin{center}
    \includegraphics[scale=0.9, angle=270]{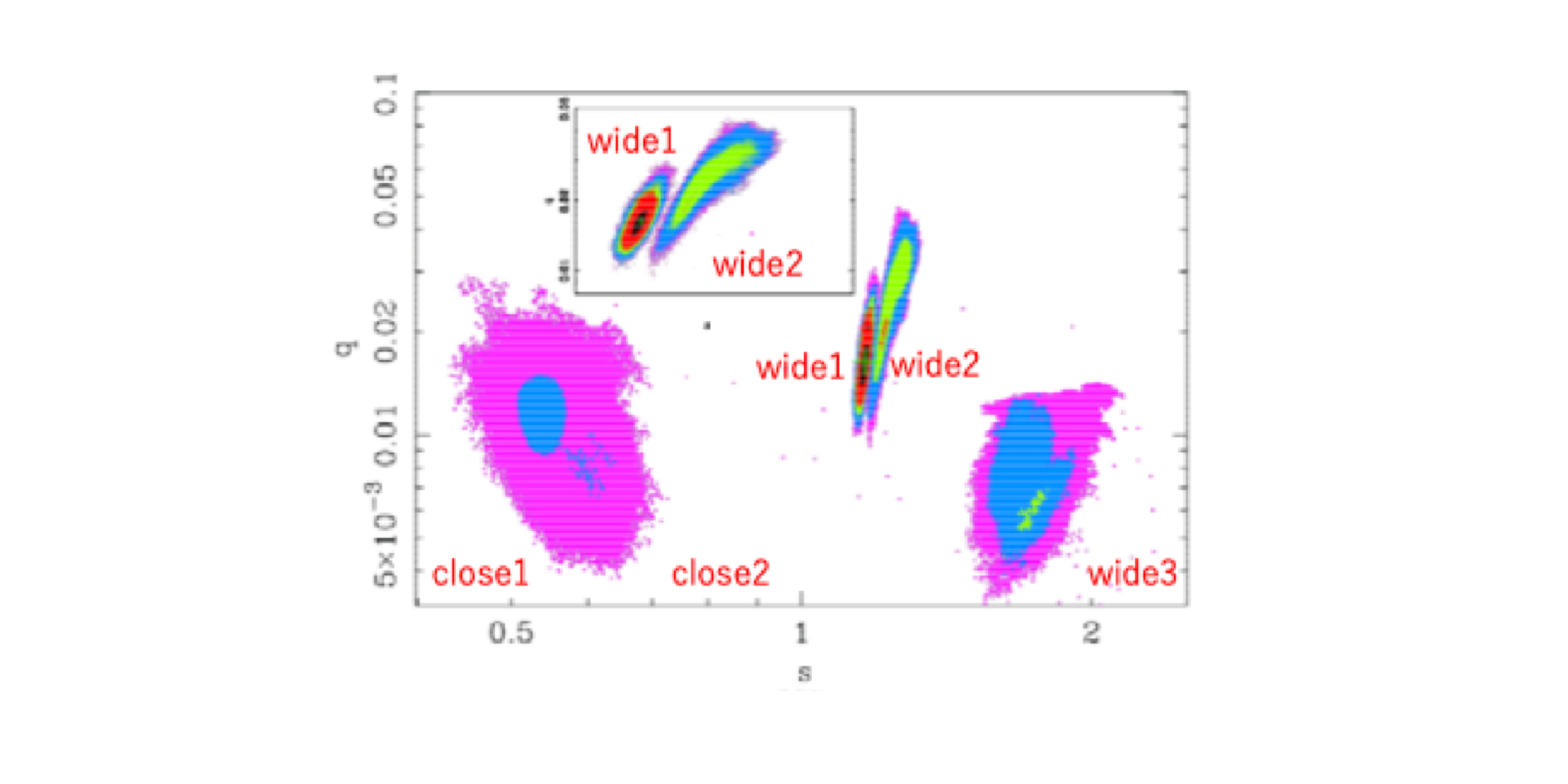}
   \caption{The distribution of the planet/host mass ratio, $q$, and planet-host separation, $s$, from the Markov chains for our five degenerate models. The points are color coded based on $\Delta \chi^2$ from the best model. The black, red, green, blue, and magenta points are chains with $\Delta\chi^2 \leq 1,6,9,16,\ {\rm and}\ 25$, respectively. The red dots ($\Delta\chi^2 \leq 6$) are shown in order to clarify the local minima for the wide2 model ($\Delta \chi^{2} \sim 5.6$). The green cross shows the smallest $\chi^2$. The inset shows a zoom in the range of $8.0\times10^{ -3 } \leq q \leq 5.0\times10^{ -2 }$ and $1.1 \leq s \leq 1.4$.}
  \label{fig-mcmc}
\end{center}
\end{figure}

\begin{figure}[h]
 \begin{minipage}{0.5\hsize}
 	\centering
	\includegraphics[scale=0.36,angle=270]{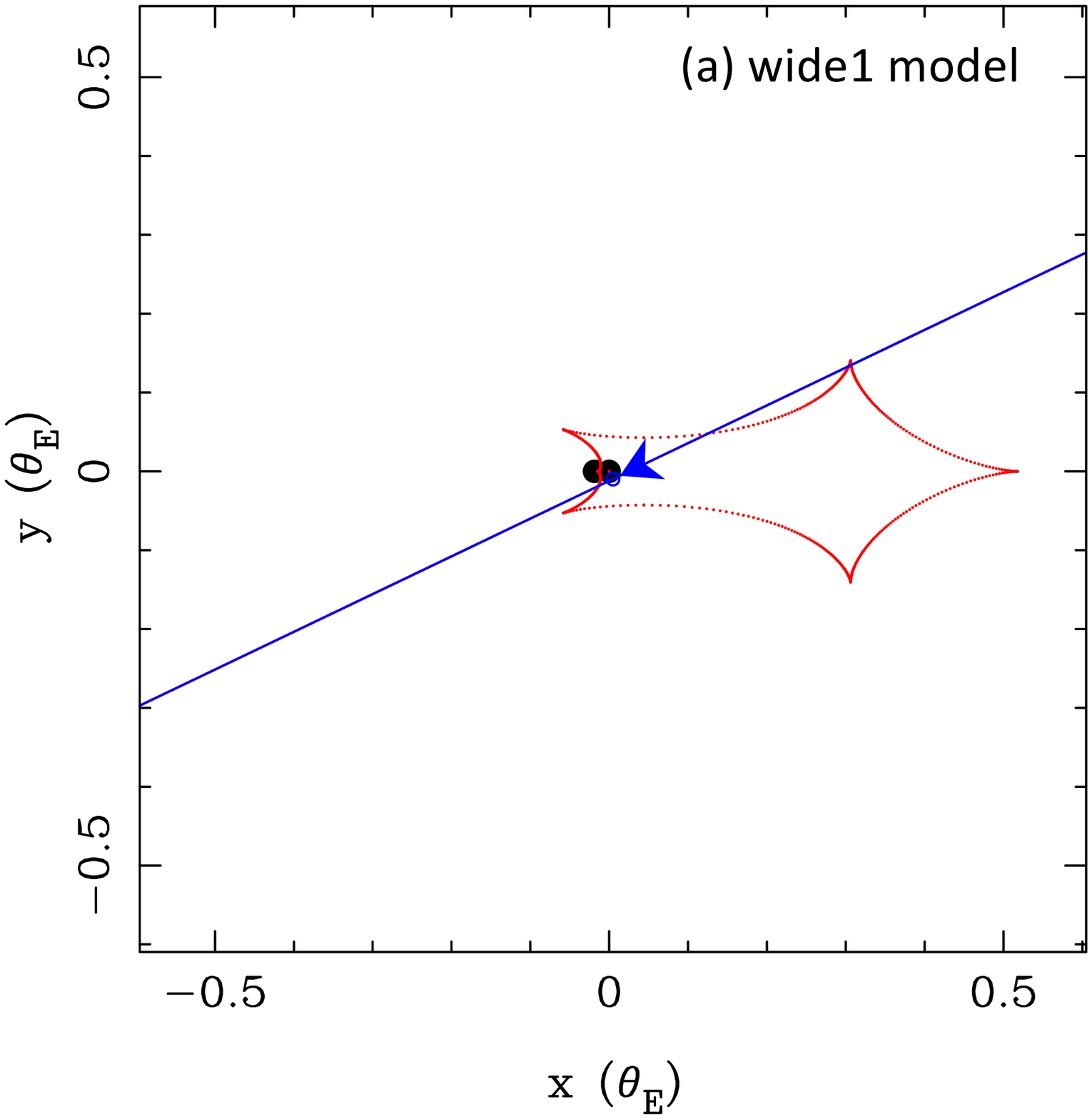}
\end{minipage}
\begin{minipage}{0.5\hsize}
 	\centering
	\includegraphics[scale=0.36,angle=270]{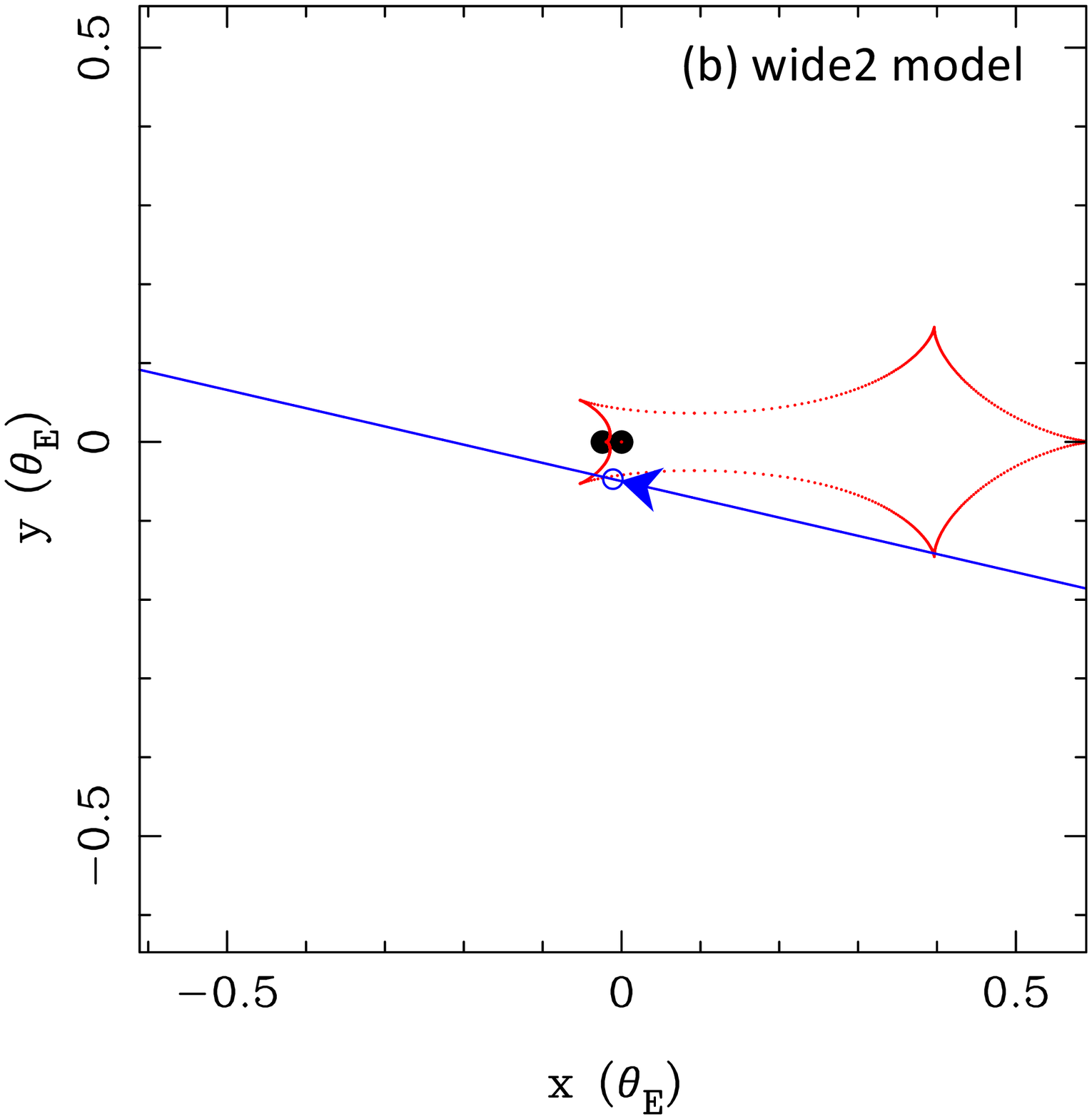}
\end{minipage}
 \begin{minipage}{0.5\hsize}
 	\centering
    \includegraphics[scale=0.36,angle=270]{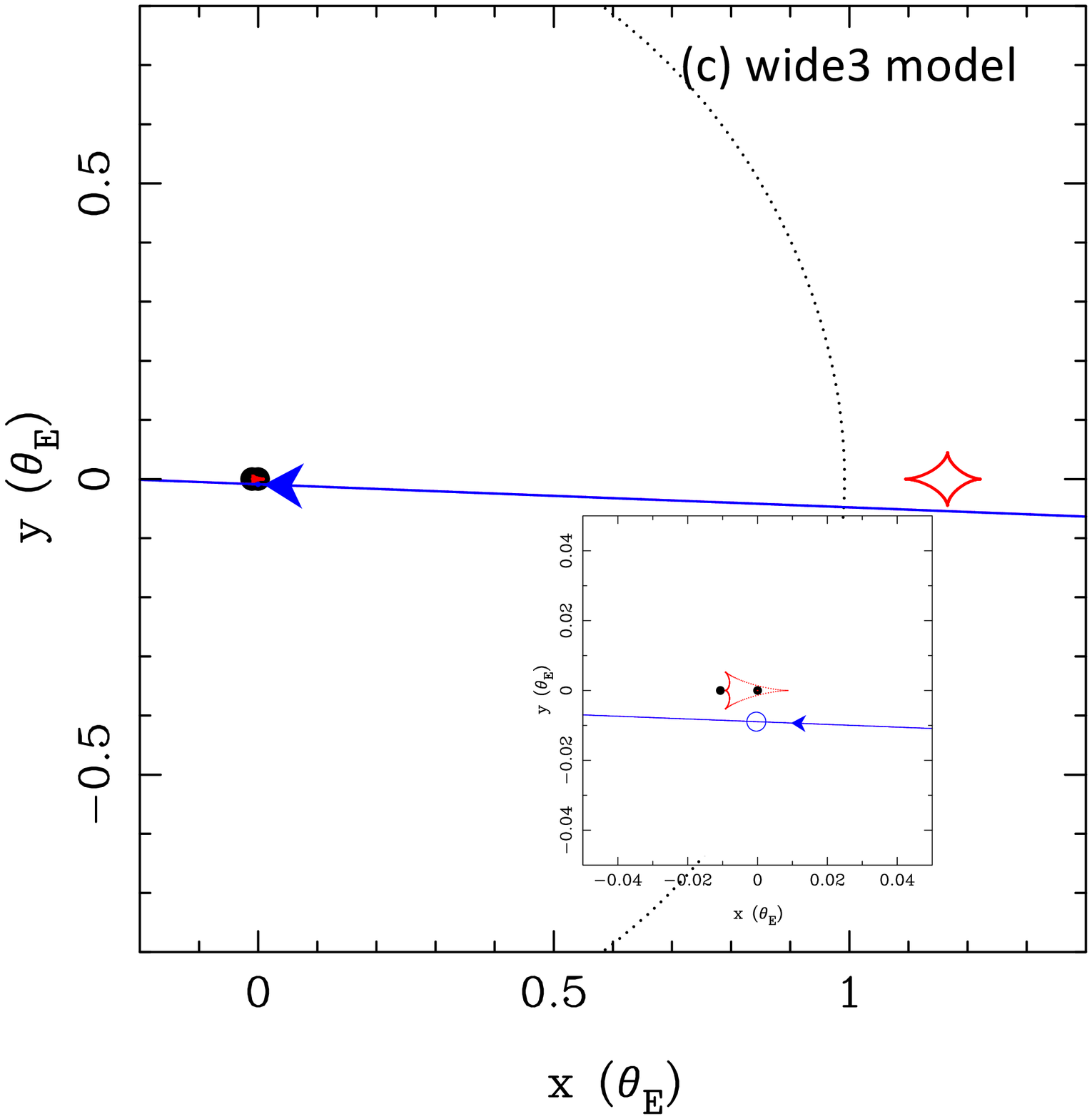}
\end{minipage}
\begin{minipage}{0.5\hsize}
 	\centering
	\includegraphics[scale=0.36,angle=270]{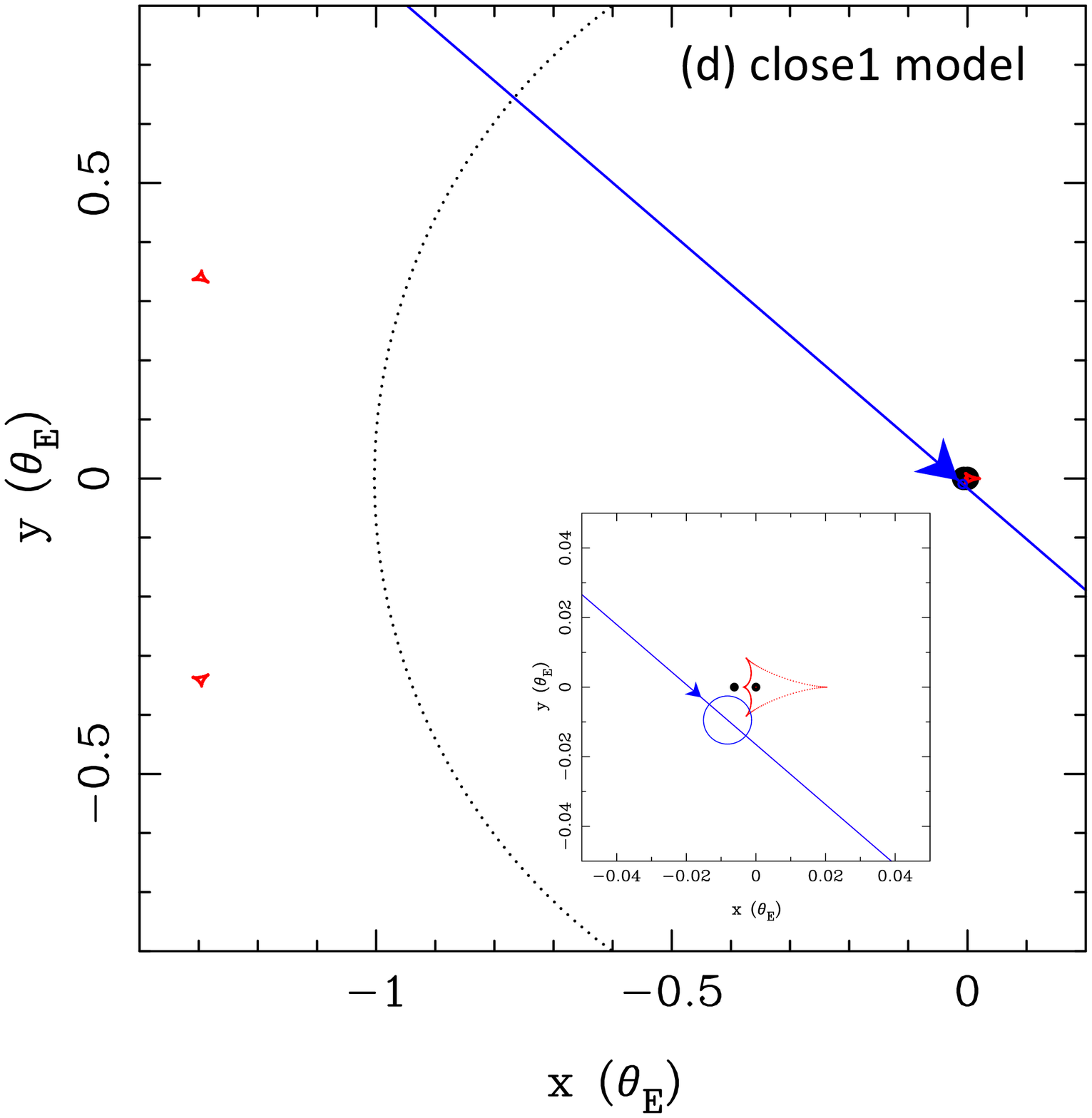}
\end{minipage}
 \begin{minipage}{0.5\hsize}
 	\centering
	\includegraphics[scale=0.36,angle=270]{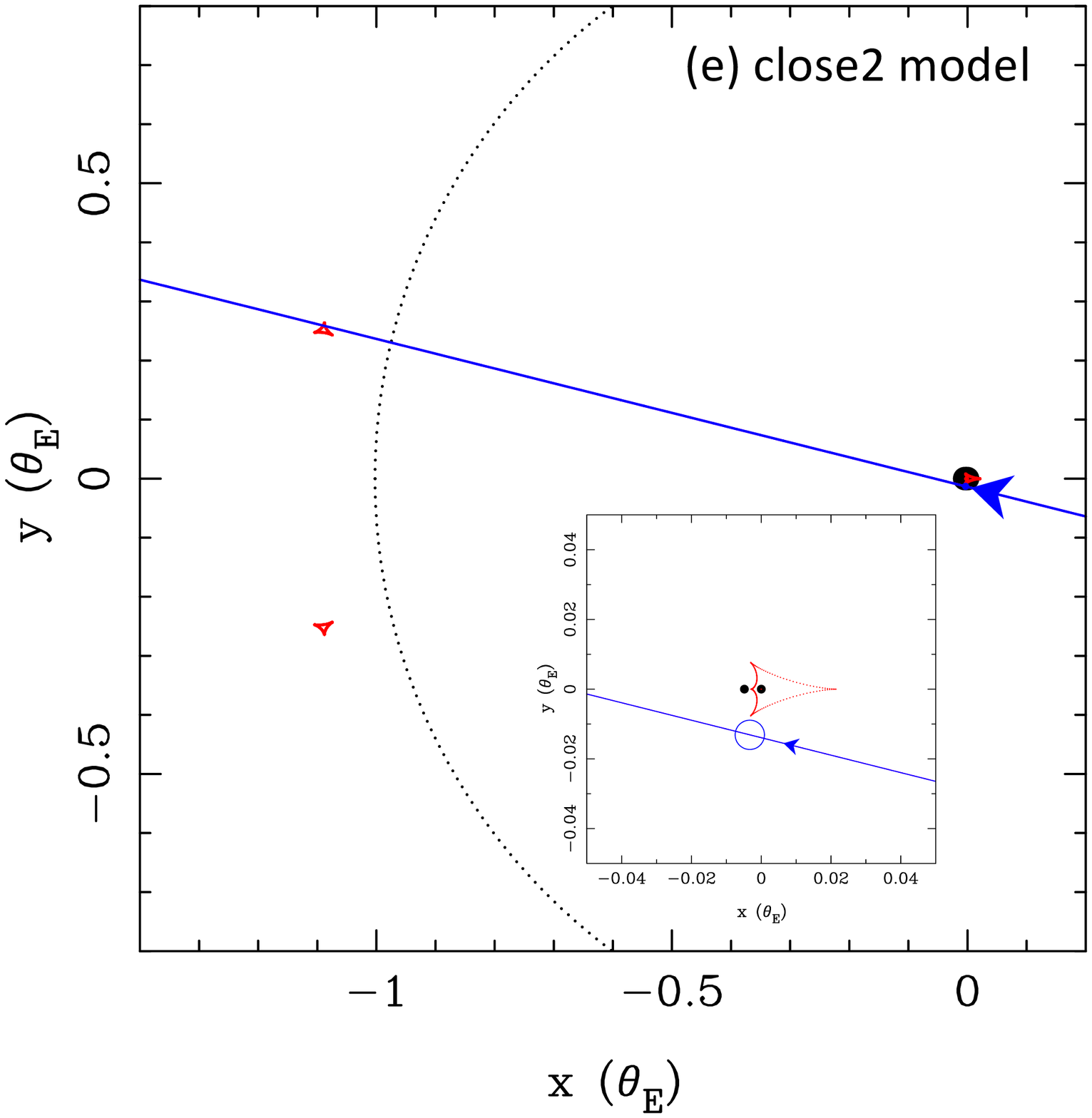}
\end{minipage}
  \caption{Caustic topologies of five competing models (red lines). The blue lines show the source trajectories on the lens plane and the arrows indicate the direction of the source/lens relative proper motion. The blue open circles indicate the source size. The black dotted lines show the critical curves. As for the wide3, close1, and close2 models, each inset shows a zoom around the central caustic.}
  \label{fig-cau}
 \end{figure}

\begin{figure}[h]
\begin{center}
    \includegraphics[scale=0.5, angle=270]{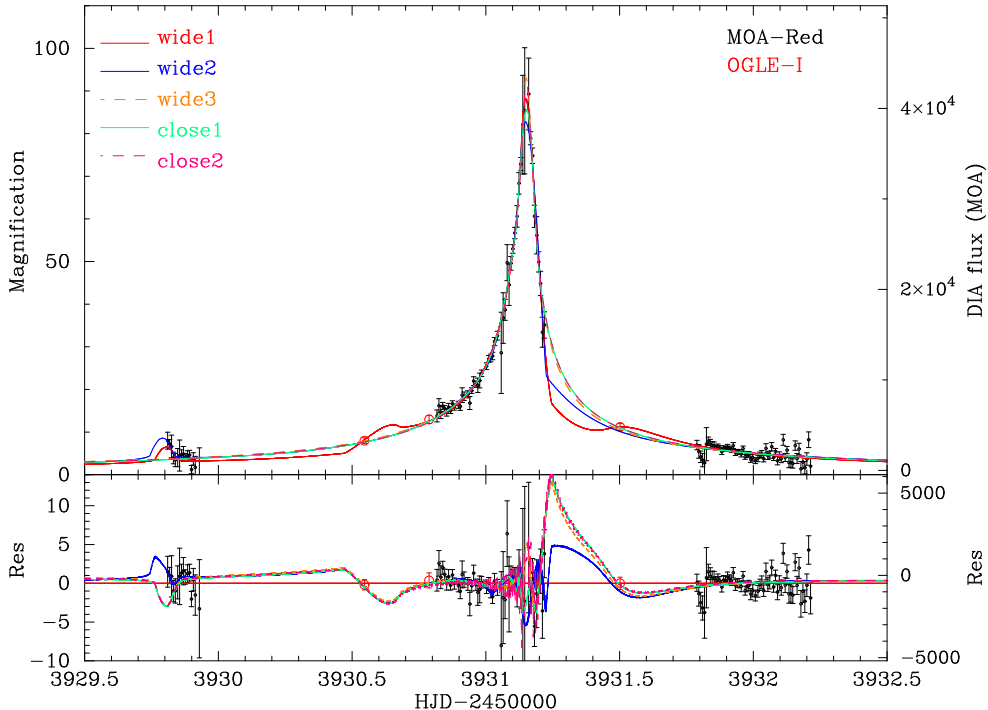} 
   \caption{The top panel shows the light curve of five competing planetary models for MOA-bin-29.  The vertical axis on the right shows the DIA flux of MOA, and the left vertical axis shows the magnification. The red line shows the best-fit model, the wide1 model;  the blue line shows the wide2 model ($\Delta \chi^{2} \sim 5.6$); the orange dashed line shows the wide3 model ($\Delta \chi^{2} \sim 7.1$); the green line shows the close1 model ($\Delta \chi^{2} \sim 13.2$); and the pink dashed line shows the close2 model ($\Delta \chi^{2} \sim 13.9$). The bottom panel shows the residuals from the wide1 model, the best-fit model. According to this figure, the light curves of the wide1 model and the wide2 models have  similar features, such as the bump during $\rm HJD'$ $\sim$ $3929.7-3929.9$. The light curves of the two close models are also similar. In Figure \ref{fig-light2}, we take a closer look in order to clarify the difference among the five competing models.}
   \label{fig-light1}
\end{center}
\end{figure}

\begin{figure}[h]
\begin{minipage}{0.5\hsize}
 	\centering
	\includegraphics[scale=0.37,angle=270]{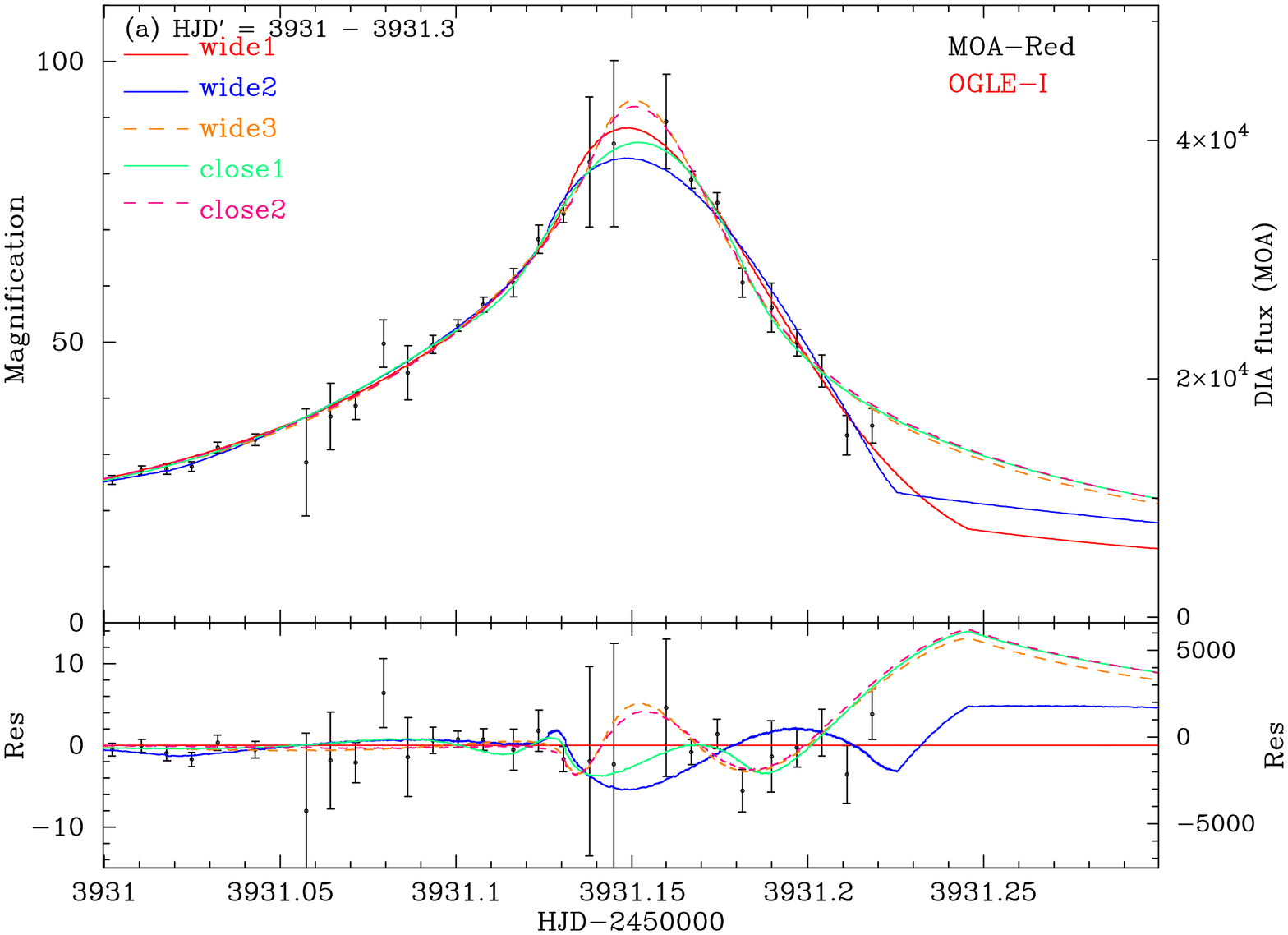}
	\label{fig-light2-1}
\end{minipage}
\begin{minipage}{0.5\hsize}
 	\centering
	\includegraphics[scale=0.37,angle=270]{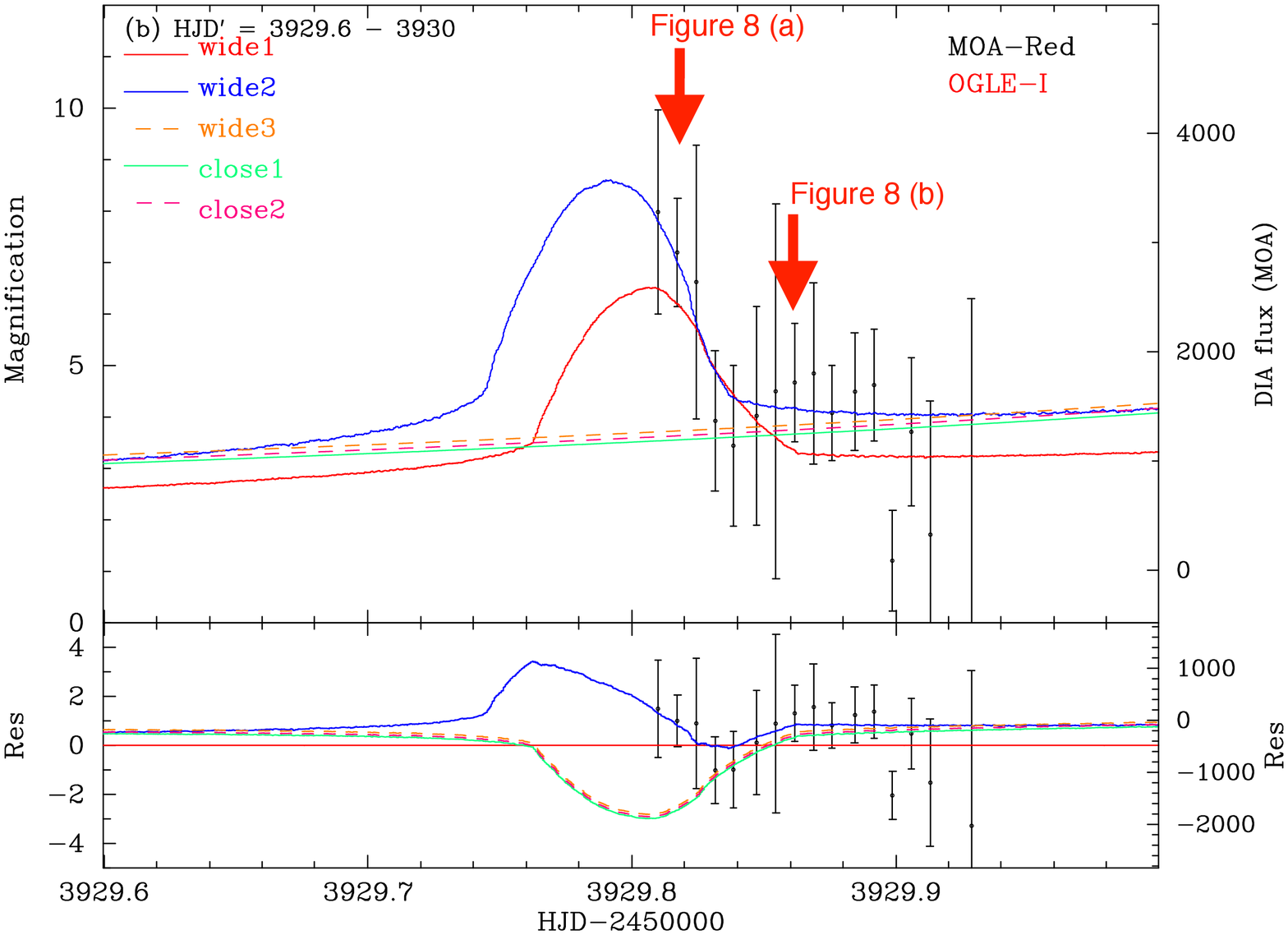}
	\label{fig-light2-2}
\end{minipage}
\\ \\
\begin{minipage}{0.5\hsize}
 	\centering
	\includegraphics[scale=0.37,angle=270]{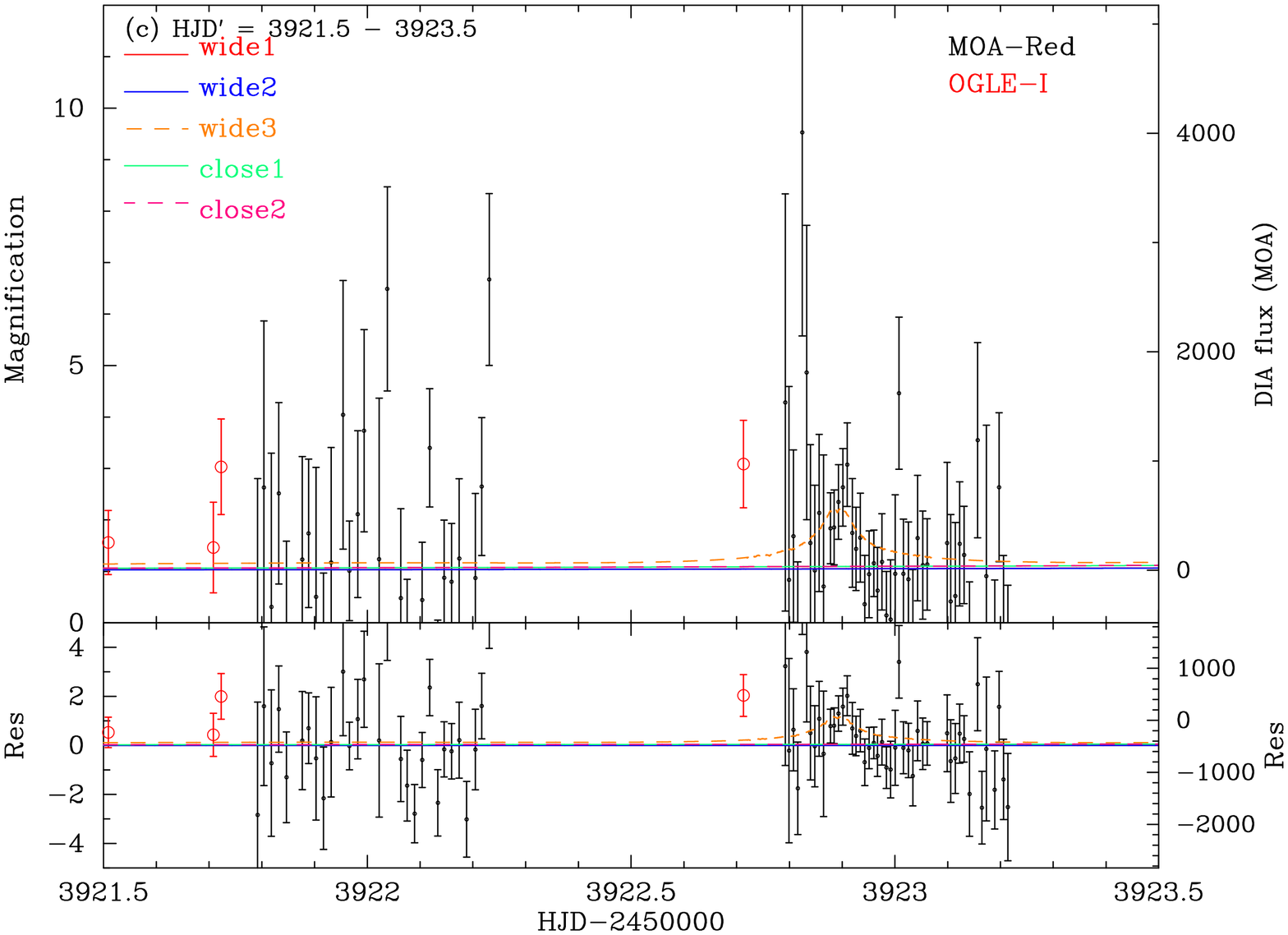}
	\label{fig-light2-3}
\end{minipage}
\begin{minipage}{0.5\hsize}
 	\centering
	\includegraphics[scale=0.37,angle=270]{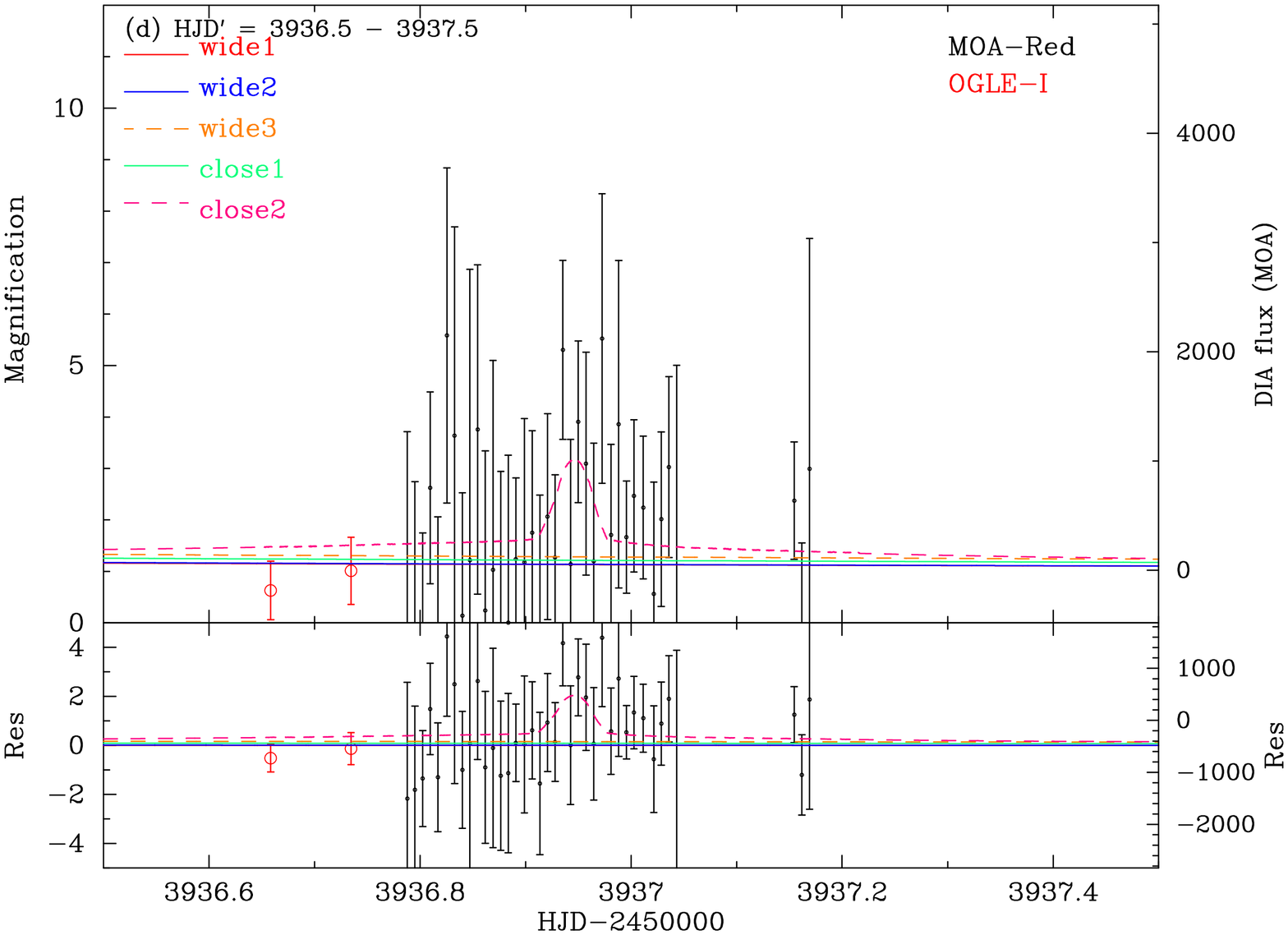}
	\label{fig-light2-4}
\end{minipage}
  \caption{Close-ups showing the five competing planetary models of the MOA-bin-29 light-curve at times of interesting light curve features (top panels). The light curves for each model are shown with the same color scheme as Figure \ref{fig-light1}, the best-fit model, the wide1 model (red), the wide2 model ($\Delta \chi^{2} \sim 5.6$) (blue), the wide3 model ($\Delta \chi^{2} \sim 7.1$) (orange), the close1 model ($\Delta \chi^{2} \sim 13.2$) (green), and the close2 model ($\Delta \chi^{2} \sim 13.9$) (pink). The bottom panels show the residuals from the wide1 model. 
We cannot conclude if these small bumps in (b)--(d) were due to the systematics or real features on the baseline.  However, we concluded that they are too insignificant to affect to the final result.}
  \label{fig-light2}
 \end{figure}

By the grid search, we found the best binary-lens model and  some local minima. The best binary model, wide1 model, is favored over the single-lens model by $\Delta \chi^{2} \sim 154$. Figure \ref{fig-pspl} shows a clear anomaly in the light curve from the single-lens model, and the best-fit binary-lens model can explain the anomaly near the peak. Figure \ref{fig-mcmc} shows locations of these degenerate models in $q-s$ space from our five Markov chains of the $\chi^{2}$ distribution of the planet/host mass ratio $q$, and the planet-host separation, $s$, in the range of $4.0\times10^{ -3 } \leq q \leq 1.0\times10^{ -1 }$ and $0.4 \leq s \leq 2.4$. The points are color coded based on $\Delta \chi^{2} $ from the best $\chi^{2}$ of the best model, the wide1 model.
Thus, we took a closer look at these models.
 Figure \ref{fig-cau} shows the caustic geometries of each model, and Figure \ref{fig-light1} shows the light curves of all degenerate models. The parameters of models are listed in Table \ref{tab-para}. Figures \ref{fig-light2} shows zooms of the characteristic part of the light curves.
\\

\begin{table}[h]
\begin{threeparttable}
\caption{The Best-fit parameters for five competing models}
\label{tab-para}
 \begin{tabular}{cc|rrrrrrrr}
 \hline \hline
 Parameters     	  &  Unit                &  wide1 & wide2 & wide3 &  close1 & close2                                          \\  \hline 
 $t_{0}$      	            &  HJD-2450000         &  3931.097 & 3931.098 & 3931.080 & 3931.135  & 3931.139  \\         
                 		 &                      &  0.003 & 0.026 & 0.009 & 0.002 &  0.004                     \\         
 $t_{\rm E}$   	   &  days                &  3.887 & 3.139 & 7.014 & 5.183  & 5.191                      \\          
                  &                      &  0.210 & 0.150 & 0.920 & 0.860 & 0.950                                   \\         
    $u_{0}$          &  $\times10^{ -2 }$  &  1.090 & 4.849 & 0.893  & 1.249 &  1.351                       \\         
                  &                      &  0.076 & 0.690 & 0.290  & 0.320  & 0.320                                 \\         
 $q$              &  $\times10^{ -2 }$  &  1.631 & 2.038 & 0.614  & 1.170 &  0.827                          \\         
                  &                      &  0.240 & 0.630 & 0.210  & 0.240 & 0.210                                  \\         
 $s$              &                      &  1.164 & 1.219 & 1.745  & 0.537  & 0.589                          \\         
                  &                      &  0.009 & 0.030 & 0.097  & 0.024  & 0.032                           \\         
    $\alpha$         &  radian              &  2.695 & 3.369 & 3.103 & 3.852  & 3.387                      \\ 
                  &                      &  0.027 & 0.073 & 0.048  & 0.04  & 0.076                           \\         
 $\rho$           &  $\times10^{ -2 }$  &  0.803 & 1.233 & 0.271  & 0.691  & 0.420                          \\         
                 &                      &   0.092 & 0.120 & 0.097  & $(< 0.871)$\tnote{a} & $(<0.540)$\tnote{a}                          \\  \hline 
 $\Delta \chi^{2}$  &                      &   & 5.55 & 7.08 & 13.24  & 13.89  \\  
 \hline \hline
\end{tabular}
\begin{tablenotes}
\item[a] This value indicates a $1\sigma$ upper limit on $\rho$. The close1 and the close2 models are favored by only $\Delta \chi^{2} \sim 17$ and $\sim 6$, respectively, over models with $\rho=0$. Because of the weak measurements of $\rho$ for the close models, we put upper limits on $\rho$. 
\end{tablenotes}
\end{threeparttable}
\end{table}


The five competing models are divided into wide models, with $s\ >\ 1$, and close models, with $s\ <\ 1$. The Einstein radii crossing times are short ($t_{\rm{E}}$$\sim 4-7$ days) for all models, and all models have a planetary mass ratio.
\\

We now describe each of these degenerate models.
\\
The wide1 model: This is the best-fit model with a planetary mass ratio of $q = 1.6 \times10^{ -2 }$ and a separation of $s = 1.2$. Figure \ref{fig-cau} (a) and \ref{fig-light1} shows the light curve and the caustic of this model.
The Einstein radius crossing time $t_{\rm E}$ is only 4 days.
We can see a bump around around $\rm HJD'$ $\sim$ $3929.7-3930.0$ due to cusp crossing in Figure  \ref{fig-light2} (b) followed by the main peak due to the caustic exit where the MOA data have a good coverage in Figure \ref{fig-light2} (a). Thanks to its caustic crossing feature, the clear finite source effects are detected by $\Delta \chi^{2} \sim 64$. 
\\

The wide2 model: Because the parameters such as $q$ and $s$ are slightly different from those of the wide1 model, the shape of the caustic is similar to that of the wide1 model in Figure \ref{fig-cau} (b). We can see a bump similar to as that of the wide1 model on $\rm HJD'$ $\sim$ $3929.7-3930.0$ because the source crosses the similar cusps as shown in Figure \ref{fig-light2} (b). Although the source crosses a different part of the caustic from that of model 1, the features of the light curves of both models are alike during the data coverage.
\\

The wide3 model: This model has a clearly different features in the caustic shape  and the light curve from those of the wide1 model. This model is disfavored against the model 1 by $\Delta\chi^2 \sim 7.1$.
The mass ratio is $q = 0.6 \times10^{ -2 }$ and the separation is $s = 1.7$, which is larger than that of the wide1 model. Additionally, the Einstein radius crossing time is about 7 days, which is twice as long as that of the wide1 model. From the light curve in Figure \ref{fig-light2} (c), we find another bump around $\rm HJD'$ $\sim$ $3921.5-3923.5$ due to a cusp approach to a planetary caustic, and the main magnification arises by approaching to a cusp of the central caustic.
\\

The close1 and close2 model: These two models have star--planet separations of $s < 1$. The close1 and the close2 models are disfavored by $\Delta\chi^2 \sim 13.2$ and $\Delta\chi^2 \sim 13.9$, respectively.
These two models have similar parameters, so the shapes of caustic geometry of both models are similar in Figure \ref{fig-cau} (d) and (e). As for the close1 model, the mass ratio is  $q = 1.2 \times10^{ -2 }$ and the separation is $s = 0.5$, which is smaller than that of the wide1 model. The Einstein radius crossing time is about 5 days. 
The light curves of both models are characterized by a cusp approach of the central caustic.
The difference between the models is the source trajectory, and regarding close2 model, a bump around $\rm HJD'$ $\sim$ $3936.8-3937.2$ comes from a planetary caustic crossing (Figure \ref{fig-light2} (d)).
\\

According to Figure \ref{fig-light1} and \ref{fig-light2}, 
the difference between models are characterized by some bumps. The wide1 and the wide2 models have a bump at $\rm HJD'$ $\sim$ $3929.7-3930.0$, and the wide3 model has a bump at $\rm HJD'$ $\sim$ $3921.5-3923.5$. The close1 model does not have a bump, but the close2 model has a bump at $\rm HJD'$ $\sim$ $3936.8-3937.2$.
\\

\begin{figure}[h]
\begin{minipage}{0.5\hsize}
 	\centering
	\includegraphics[scale=0.30, angle=90]{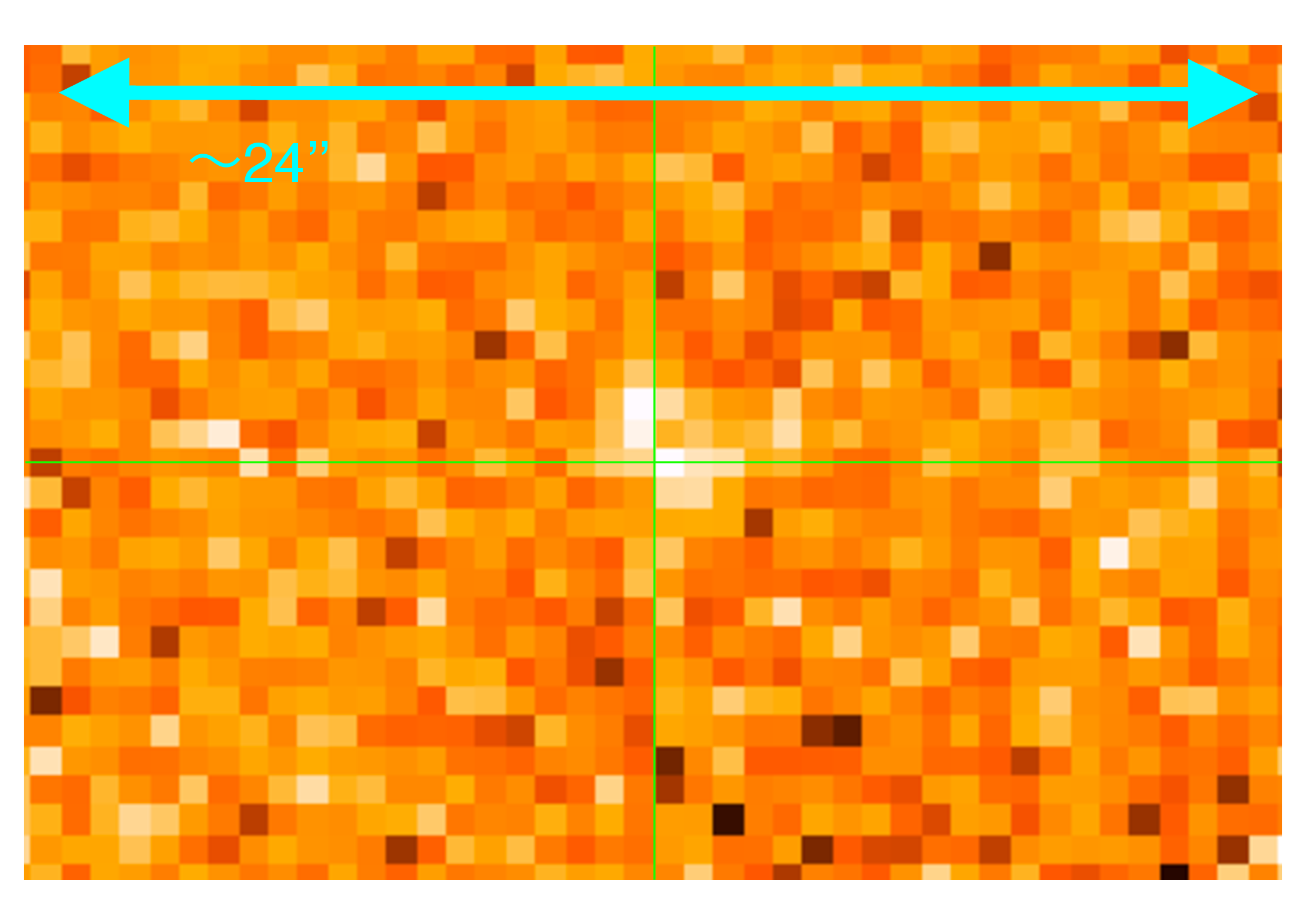}
	\hspace{1.6cm} (a)\ at\ $\rm HJD'$\ $\sim$\ $3929.82$
\end{minipage}
\begin{minipage}{0.5\hsize}
 	\centering
	\includegraphics[scale=0.30, angle=90]{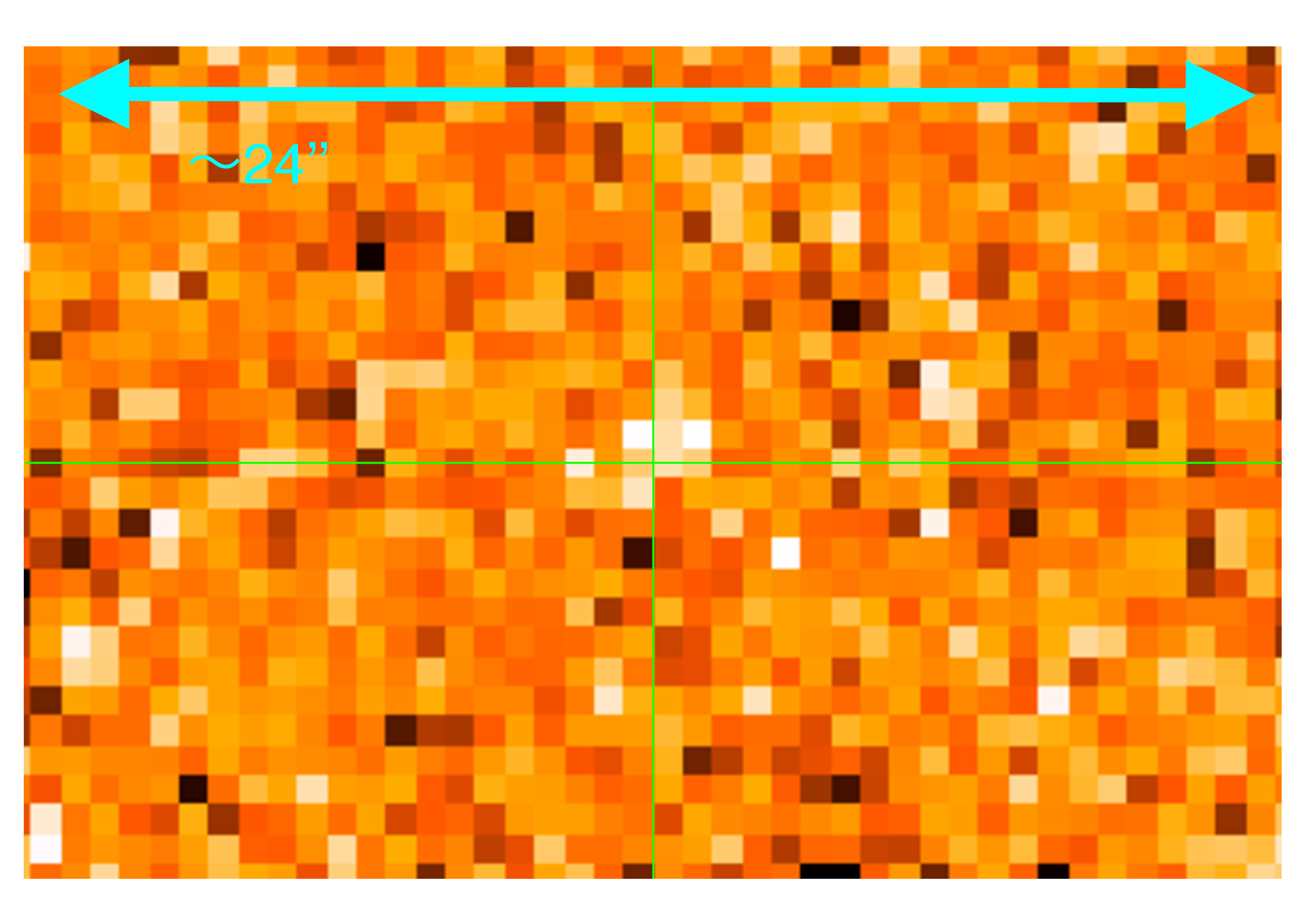}
	\hspace{1.6cm} (b)\ at\ $\rm HJD'$\ $\sim$\ $3929.88$
\end{minipage}
\caption{\bf The difference image at $\rm HJD'$ $\sim$ $3929.82$ and at $\rm HJD'$ $\sim$ $3929.88$. The green cross shows the position of the event. North is up and east is to the left. This difference image indicates a variable object to be detected. The variable object in the difference image at $\rm HJD'$ $\sim$ $3929.82$ is brighter than any of that in the different images during $\rm HJD'$ $\sim$ $3929.82-3929.94$. The magnification at $\rm HJD'$ $\sim$ $3929.82$ is about 7, while that at $\rm HJD'$ $\sim$ $3929.88$ is about 4. They are marked with arrows in Figure \ref{fig-light2} (b).}
   \label{diff}
 \end{figure}

The bump around $\rm HJD'$ $\sim$ $3929.7-3930.0$ is most likely to be the real caustic feature for several reasons from the viewpoint of the data. 
First, the weather was clear at night during $\rm HJD'$ $\sim$ $3929.80-3929.85$ and the seeing was also good. Around $\rm HJD'$ $\sim$ $3929.80$,
the flux ($\sim 3000$) was significantly larger than those of other small bumps ($\sim 1000$).
Second, we checked the difference images during the bump and found that the difference images at $\rm HJD'$ $\sim$ $3929.82$
indicate a variable object to be detected (Figure \ref{diff} (a)). The center of magnification of that image is consistent with those of the difference images around the main peak at $\rm HJD'$ $\sim$ $3931$.
Third, we found that the variable object in the difference image at $\rm HJD'$ $\sim$ $3929.82$ is brighter than that of any in the different images during $\rm HJD'$ $\sim$ $3929.82-3929.94$ (one of the images is in Figure \ref{diff} (b)).
Therefore, the bump around $\rm HJD'$ $\sim$ $3929.7-3930.0$ might have been caused by astrophysical origins.
\\

However, it is difficult to judge if other small bumps (around $\rm HJD'$ $\sim$ $3921.5-3923.5$ and $\rm HJD'$ $\sim$ $3936.8-3937.2$) are due to the systematics on the baseline or the real caustic feature because these data points do not indicate any bad seeing, air masses, nor $\chi^2$ in PSF fitting. 
The $\chi^2$ differences due to these bumps are small, $\sim 10$ at most. We can see a lot of $2-3\sigma$ outliers that can cause similar bumps on the baseline data. 
The visual inspection of difference images during the bumps do not indicate any strange features such as saturation images, 
but a significant variable object cannot be detected in them.
Although the effects of the airmass, the seeing, and the differential refraction for each data point have been corrected during the data reduction, it is not surprising that such small systematics remain in that dense stellar field. 
\\

Thus, we investigated whether the existence of these bumps 
affects the final results in the Appendix \ref{app-ba}. 
As a result, we found that the final conclusion of the analysis remains the same whether we include these bumps or not, because they do not affect the MCMC distributions.
However, the small $\chi^2$ differences cannot be used to compare the competing models. Instead, the range of parameters in these models, $6.0\times10^{ -3 } \leq q \leq 2.0\times10^{ -2 }$ and $0.5 \leq s \leq 1.8$, should be taken as an uncertainty, conservatively. Therefore, we use these original five models without removing any the data points but recognize them as best-fit models equally, not weighting by $\Delta \chi^2$,  in a Bayesian analysis in Section \ref{sec-bay}.

We also checked if the caustic feature producing the bump at
$\rm HJD'$ $\sim$ $3929.7-3930.0$ is real or not from the viewpoint of modeling, and reached two conclusions. 
First, we found new local minima with a similar bump around $\rm HJD'$ $\sim$ $3929.7-3930.0$ when we 
conducted a new light-curve modeling grid search after removing all of the data points from that night. 
This implies that the bump is a real caustic feature. 
Second, this bump is closer to the center of mass, which is more likely to cross the caustics.
On the other hand, the other bumps due to the planetary caustic are much farther from the center of mass, which means that the source is much more likely to miss these small caustics, 
particularly if planetary orbital motion or microlensing parallax was included in the modeling. 
So, the rest of the light-curve data implies that the bump at $\rm HJD'$ $\sim$ $3929.7-3930.0$ is real, but it does not imply that the other light-curve bumps are real.
Therefore, we conclude the bump at $\rm HJD'$ $\sim$ $3929.7-3930.0$ is a real caustic feature.
\\

We could detect clear signals of the finite source effects in the three wide models, but we obtained very weak measurements of $\rho$ for the close models. The best-fit close models are favored by only $\Delta \chi^{2} \sim 17$ and $\sim 6$, respectively, over models with $\rho=0$. We detect strong signals of the finite source effects for the other models.
\\

\subsection{Binary-source model}
\begin{figure}[htbp]
\begin{center}
   \includegraphics[scale=0.50, angle=270]{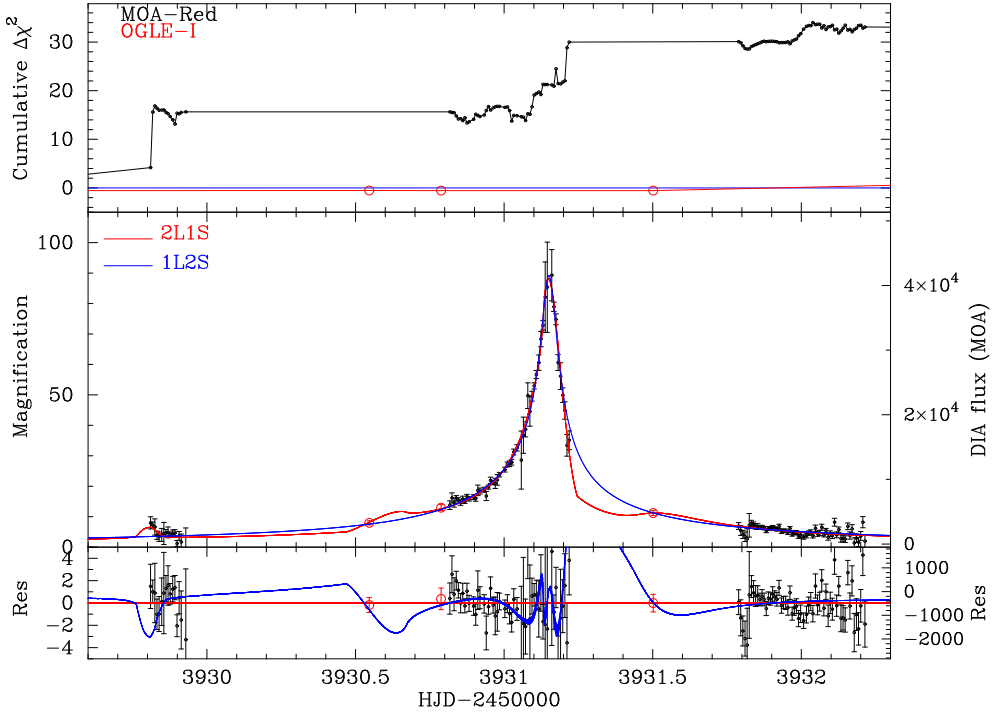}
   \caption{The top panel shows the difference in the cumulative $\chi^2$ between the 2L1S and 1L2S models. The middle and bottom panels show the light curves of 2L1S model (red) and 1L2S model (blue) and the residuals from 2L1S model, respectively.}
   \label{fig-bs}
\end{center}
\end{figure}

There is a possible degeneracy between the single-lens, binary-source model (1L2S) and the binary-lens, single-source model (2L1S)  \citep{1993ApJ...407..440G}. 
For the 1L2S model, the magnification $A$ is expressed in the following equation:
\begin{equation}
  A = \frac{A_{1}F_{1} + A_{2}F_{2}}{F_{1} + F_{2}} = \frac{A_{1} + q_{F}A_{2}}{1 + q_{F}},
\end{equation}
 where $A_{1}$ and $A_{2}$ are the magnification of the two sources with flux $F_{1}$ and $F_{2}$, respectively, and $q_{F}$ is the flux ratio between the two sources ($=F_{2}/F_{1}$).      
For 1L2S model, the magnification $A$ depends on the wavelength unless the two source stars have the same color. By using the color difference expected for the two sources of unequal luminosity, the 2L1S/1L2S degeneracy could be solved \citep{1998ApJ...506..533G}. 
For some microlensing events, the 2L1S/1L2S degeneracy is broken with this method, and this confirms the planetary models \citep{2014MNRAS.439..604S, 2018AJ....156..236Z}. 
However, due to the poor data coverage for this event, the model parameters are uncertain, and this makes it difficult to use the color-shift method mentioned above to confirm the planetary interpretation more strongly.
We searched for the best 1L2S model and found that the best 1L2S model is disfavored over 2L1S model by $\Delta \chi^{2} = \chi^{2}_{\rm 2L1S} - \chi^{2}_{\rm 1L2S} \sim 29$. Figure \ref{fig-bs} shows the comparison between 2L1S/1L2S. Table \ref{tab-bs} shows the parameters of  the best-fit 1L2S model.
According to Figure \ref{fig-bs}, the sizable fraction of $\chi^2$ differences arise from the bump,
which we conclude is the real caustic feature.
The $\chi^2$ difference from the outside of the bump is $\chi^{2}_{\rm 2L1S} - \chi^{2}_{\rm 1L2S} \sim 12.84$. So, the best-fit 1L2S model is disfavored over the 2L1S model both from the bump and from the rest of the light curve. 
Thus, we conclude that the possibility of the 1L2S model is excluded.
\\

\begin{table}[h]
\caption{The Best-fit parameters and the median value with 68.3\% confidence interval derived from MCMC chains for the 2S1L model}
\begin{center}
 \begin{tabular}{cc|rrrrrrrr}
 \hline \hline
 Parameters       &  Unit                &  Best fit  & MCMC                                   \\  \hline 
 $t_{0,1}$          &  HJD-2450000         &  3931.151 & $3931.150_{-0.001}^{+0.001}$ \\         
 $t_{0,2}$          &  HJD-2450000         &   3931.090 & $3931.082_{-0.024}^{+0.026}$\\         
 $t_{\rm E}$      &  days                &     4.486  &$3.998_{-0.689}^{+0.693}$             \\         
 $u_{0,1}$          &  $\times10^{ -2 }$  &   -0.652  &  $-0.905_{-0.212}^{+0.218}$                   \\         
 $u_{0,2}$          &  $\times10^{ -2 }$  &      4.732  & $5.754_{-1.617}^{+1.668} $                \\         
 $\rho_1$           &  $\times10^{ -2 }$  &         0.287  &  $0.636_{-0.426}^{+0.442}$            \\         
 $\rho_2$           &  $\times10^{ -2 }$  &       4.788 &  $2.606_{-1.552}^{+1.546}  $            \\         
 $q_{F, {\rm OGLE}} $   &        &  1.309  & $12.351_{-10.566}^{+13.266}$ \\
 $q_{F,{\rm MOA}} $   &         &    1.004  &  $0.911_{-0.175}^{+0.165}$    \\   \hline
 \hline
\end{tabular}
\label{tab-bs}
\end{center}
\end{table}

\subsection{Microlensing parallax model}
The microlensing parallax is an effect caused by the orbital motion of the Earth. 
Although it is known that there is little possibility of the detection of the parallax effect for such a short duration event \citep{1992ApJ...396..104G, 2012ARA&A..50..411G}, we also considered a parallax model for completeness. Then, we found that the parallax model improves the fit only by  $\Delta \chi^{2} \sim 8.40$, and the value of the parameters are $\pi_{\rm E,E}=407\pm95$ and $\pi_{\rm E,N}=352\pm88$, which are quite larger than the ordinary value ($<\ 1$). Therefore, we ruled out the parallax model.\\

\section{Angular Einstein Radius}
\label{sec-cmd}
\begin{figure}[h]
\begin{center}
   \includegraphics[scale=0.45, angle=270]{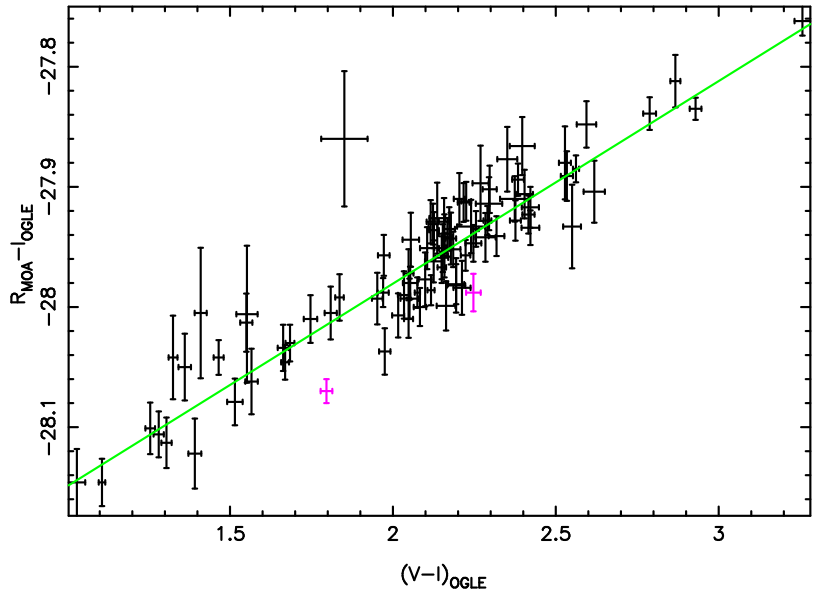}
   \caption{Color--color relation between $(R_{\rm MOA} - I_{\rm OGLE})$ and $(V - I)_{\rm OGLE}$. In order to derive this formula, we cross-reference stars in the MOA DoPHOT catalog of the reference image with the stars in OGLE-III photometry map \citep{2011AcA....61...83S} within $120^{\prime\prime}$ around the source star and then use 91 cross-referenced stars. The rejected $3\sigma$ are displayed with pink crosses.}
\end{center}
\label{fig-app1}
\end{figure}

Thanks to the detection of the finite source effects, we can constrain the lens physical properties by estimating the angular Einstein radius $\theta_{\rm E}=\theta_{\rm *}/\rho$.
We can get $\rho$ from the light-curve modeling and the angular source radius $\theta_{\rm *}$ 
by using empirical relation of $\theta_{\rm *}$, the intrinsic source color $(V-I)_{\rm S,0}$ and magnitude $I_{\rm S,0}$ \citep{2014AJ....147...47B}.
Because there is no $V$-band data during the magnification, we estimated the source color as follows.
We cross-referenced stars in the MOA DoPHOT catalog of the reference image with the stars in OGLE-III photometry map \citep{2011AcA....61...83S} within $120^{\prime\prime}$ around the source star. By using 91 cross-referenced stars, we derived the following color-color relation (see Figure \ref{fig-app1}):
\begin{equation}
R_{\rm MOA} - I_{\rm OGLE} = (-28.32 \pm 0.01) + (0.16 \pm 0.01)(V - I)_{\rm OGLE}.
\end{equation}
If we have a good measurement of the OGLE $I$-band source magnitude, $I_{\rm S, OGLE}$, we could derive the $(V-I)_{\rm S}$ from this formula. However, $I_{\rm S, OGLE}$ from the light-curve fitting has very large uncertainty because only a few data points during the low magnification are available. 
\\

\begin{figure}[h]
\begin{center}
   \includegraphics[scale=0.6, angle=270]{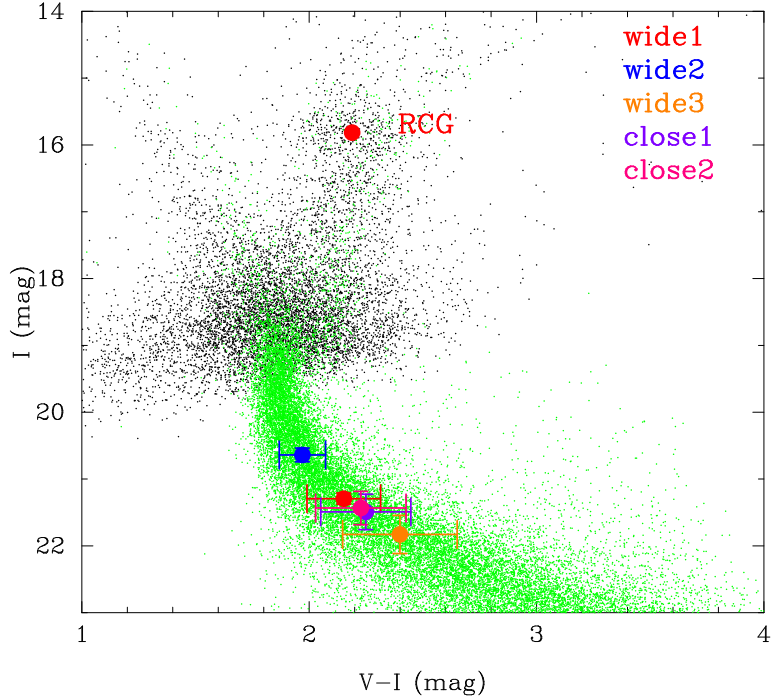}
   \caption{Color-magnitude diagram (CMD) of the stars in the OGLE-III catalog within $120^{\prime\prime}$ of the source star is shown as black dots, and the $HST$ CMD of \citep{1998AJ....115.1946H}, which is transformed to the same reddening and extinction as the same field as the event, is shown as green dots. The red dot shows the centroid of the red clump giant distribution. The colors and magnitudes of each of the five competing models are shown with red, blue, orange, purple, and pink symbols, respectively.}
\end{center}
\label{figure-cmdfinal}
\end{figure}

Therefore, by following \citet{2008ApJ...684..663B}, we estimated the source color by taking the average color of main-sequence stars in Baade's window observed by the $Hubble\ Space\ Telescope$, ($HST$, \citet{1998AJ....115.1946H}). 
In the color-magnitude diagram (CMD) (Figure \ref{figure-cmdfinal}), the black dots show the OGLE stars within $120^{\prime\prime}$ around the source star, and the green dots show the $HST$ stars that are adjusted for the reddening and extinction by using the the Red Clump Giants (RGC) color and magnitude, $(V-I, I)_{{\rm RGC, }HST} = (1.62, 15.15)$ \citep{2008ApJ...684..663B}. Because we do not have any good calibrated $I$-band source magnitudes, we derived its magnitude and color as follows. 
We solve for $I_{\rm S}$ and $(V-I)_{\rm S}$ using an iterative procedure. First, we estimate the initial source color,  $(V-I)_{\rm S}$ from the average color of the main-sequence stars with the input $I_{\rm S}$ value. We then determine the new $I_{\rm S}$ values from this color and the $R_{\rm MOA}$ values from the light-curve model. After a few iterations, this converges. We used the $I_{\rm S, OGLE}$ value from the light-curve model for the initial $I_{\rm S}$ value.

We derive the extinction-free magnitude and color of source so as to calculate an angular Einstein radius by following a method similar to that of \citet{2004ApJ...603..139Y}. The extinction-corrected magnitude can be determined from the magnitude and color of the centroid of the RGC feature in the CMD. In figure \ref{figure-cmdfinal}, the red point shows the centroid of RCG color and magnitude, $(V-I,I)_{\rm RCG} = (2.19, 15.82) \pm (0.01, 0.02)$ around the target. Assuming that the source star suffers the same reddening and extinction as the RGCs, we compare these values to the expected extinction-free RCG color and magnitude at this field of $(V-I,I)_{\rm RCG,0} = (1.06, 14.41) \pm (0.07, 0.04)$ \citep{2013A&A...549A.147B, 2013ApJ...769...88N}, and as a result, we obtained the reddening and extinction by using the RCG color and magnitude, $(V-I, I)_{{\rm RGC, }HST} = (1.62, 15.15)$ \citep{2008ApJ...684..663B}. 
\\

We determined the angular Einstein radius for each model with the following method that we demonstrate with the parameters of the wide1 model.
Assuming that the source star suffers the same reddening and extinction as the RGCs, we compare these values to the expected extinction-free RCG color and magnitude at this field of $(V-I,I)_{\rm RCG,0} = (1.06, 14.41) \pm (0.07, 0.04)$ \citep{2013A&A...549A.147B, 2013ApJ...769...88N}, and as a result, we get the reddening and extinction to the source of $(E(V-I),A_{I})_{\rm RCG,0} = (1.13, 1.40) \pm (0.07, 0.05)$ for the best-fit model, wide1 model. 
Therefore, we determined the intrinsic source color and magnitude to be
\begin{equation}
  (V-I,I)_{\rm S,0} = (1.02, 19.89) \pm (0.18, 0.10).
\end{equation}
Then, we find the angular source radius with the empirical formula \citep{2014AJ....147...47B}, 
\begin{equation}
 \log(\theta_{\rm LD}) = 0.501414 + 0.419685(V-I) - 0.2I,\end{equation}
where $\theta_{\rm LD} \equiv 2\theta_*$ is the limb-darkened stellar angular diameter, \citep{2015ApJ...809...74F}. This relation is derived by using stars with colors corresponding to $3900\ <\ T_{\rm eff}\ <\ 7000$ \citep{2017AJ....154...68B}.
We found the angular source radius $\theta_{\rm *} = 0.45 \pm 0.08 \ {\rm \mu as}$ 
for the wide1 model, with the uncertainty dominated by the source color uncertainty rather than the $2 \%$ uncertainly from the empirical formula.

\begin{table}[h]
\caption{The intrinsic Source Color, intrinsic Magnitude, Angular source Radius,  Angular Einstein Radius, and the lens-source relative proper motion}
\begin{center}
 \begin{tabular}{cc|rrrrrrrr}
 \hline \hline
 Parameters       &  Unit        &  wide1 & wide2 & wide3 & close1 & close2                                       \\  \hline 
 $I_{\rm s}$      &      mag            &   19.893 & 19.239 & 20.428 & 20.091  & 20.033  \\         
                  &                                &    0.096 &  0.112 & 0.288   & 0.269    & 0.253                    \\         
 $(V-I)_{\rm s}$  &           mag      &   1.023  & 0.842  & 1.270   & 1.120   & 1.099                    \\         
                  &                                &   0.176  &  0.112 &  0.261  &  0.210 &  0.209                                 \\    
$\theta_*$  &   $\mu$as              &   0.448 & 0.507  & 0.444   & 0.449   & 0.451                    \\         
                  &                               &   0.079  &  0.066 &  0.127  &  0.107 &  0.105                                 \\   \hline
$\theta_{\rm E}$	&	mas	&	0.056	& 0.041	&	0.164	&	  $>0.042$	 &	$>0.068$  \\
			&		&	0.012	& 0.007	&	0.075  	&	&	\\
$\mu_{\rm rel}$	&	mas ${\rm yr}^{-1}$	&	5.242	&	4.793	&	8.552	&	$>2.807$	& 	$>4.436$ \\
			&	&		1.144  &  0.823  & 4.077  &   & \\
 \hline \hline
\end{tabular}
\label{tab-mag}
\end{center}
\end{table}

Finally, we calculate the angular Einstein radius $\theta_{\rm E} = 0.056 \pm 0.012\ {\rm mas}$ and the lens-source relative proper motion $\mu_{\rm rel} = \theta_{\rm E} / t_{\rm E} = 5.242 \pm 1.144\ {\rm mas}\ {\rm yr}^{-1}$ for the wide1 model.
\\

Since the source star color and magnitude depend on the model, the angular source radius also depends on the model, and we summarize the values of $I_{\rm S, 0}, (V-I,I)_{\rm S, 0}, \theta_{\rm *}, \theta_{\rm E}$ and $\mu_{\rm rel}$ for each model in Table \ref{tab-mag}. 
As for the close models, we get only an upper limit for $\rho$, so we get only the lower limits of $\theta_{\rm E}$ and $\mu_{\rm rel}$. 
\\

\section{Lens Physical Parameters by Bayesian Analysis}
\label{sec-bay}

Because a microlensing parallax effect was not measured for this event, the lens mass cannot be directly measured from the light-curve models.
In order to estimate the probability distribution of the lens properties, we conducted a Bayesian analysis \citep{2006Natur.439..437B, 2006ApJ...644L..37G, 2008ApJ...684..663B} assuming the Galactic model of 
\citet{1995ApJ...447...53H}  as a prior probability.
We also assume a mass function used in \citet{2011Natur.473..349S} and extend it to the low-mass brown-dwarf regime ($0.001 \leq M/M_\odot$).
We use the measured $t_{\rm E}$ and $\theta_{\rm E}$ to constrain the lens physical parameters. The extinction-corrected blending flux, which includes the lens and unrelated ambient stars on the line of sight to the source star, is derived from the light-curve modeling and is set as the upper limit for the lens brightness. 
$H$-band magnitudes for the lenses of all models are estimated from the color-color relation of the main-sequence stars \citep{1995ApJS..101..117K} and the isochrone model of 5 Gyr brown dwarfs \citep{2003A&A...402..701B}. The extinctions are calculated from \citet{1989ApJ...345..245C}.
\\
\begin{table}[h]
\caption{The lens physical parameters as a result of the Bayesian analysis}
\begin{center}
 \begin{tabular}{cc|rrrrrrr}
 \hline \hline
 Parameters       &  Unit                   &  wide1 & wide2 & wide3 & close1 & close2                                         \\  \hline 
 $D_{\rm L}$      &    kpc                  &   $7.12_{-1.05}^{+1.15}$                 &  $7.23_{-1.01}^{+1.15}$         &  $6.57_{-1.31}^{+1.23}$          &  $7.07_{-1.07}^{+1.15}$    & $6.73_{-1.21}^{+1.21}$ \\           
 $M_{\rm L}$      &       $M_\sun$          &   $0.03_{-0.02}^{+0.07}$           &  $0.02_{-0.01}^{+0.06}$          &  $0.10_{-0.06}^{+0.16}$          &  $0.05_{-0.03}^{+0.08}$    & $0.07_{-0.04}^{+0.11}$  \\                 
 $M_{\rm p}$      &  $M_{\rm {Jup}}$             &   $0.60_{-0.38}^{+1.20}$   &  $0.48_{-0.31}^{+1.22}$          &   $0.63_{-0.36}^{+1.05}$   &   $0.58_{-0.37}^{+1.04}$   &  $0.58_{-0.35}^{+0.97}$    \\                  
 $r_\perp$       & au   &            $0.48_{-0.10}^{+0.12}$                         &  $0.37_{-0.07}^{+0.08}$               &  $1.54_{-0.45}^{+0.52}$  &  $0.27_{-0.07}^{+0.08}$  &  $0.39_{-0.11}^{+0.12}$  \\
 $a_{\rm 3D}$                &  au                     &   $0.59_{-0.16}^{+0.35}$                        &  $0.46_{-0.11}^{+0.27}$          &    $1.92_{-0.64}^{+1.18}$        &   $0.34_{-0.10}^{+0.21}$   & $0.49_{-0.16}^{+0.30}$\    \\ 
 $ I $               &   mag                 &       $36.35_{-8.62}^{+3.66}$               &   $37.24_{-8.57}^{+4.94}$		&  $27.77_{-2.68}^{+7.77}$	&  $35.56_{-8.54}^{+2.96}$	&  $33.59_{-7.56}^{+3.34}$   \\
$ V  $              &  mag		&  $44.61_{-11.92}^{+2.82}$		&  $45.34_{-11.60}^{+4.35}$	           &  $32.85_{-2.95}^{+11.00}$	&   $43.94_{-12.27}^{+2.23}$	&  $41.62_{-11.35}^{+3.35}$   \\
$ H $               &  mag                    &   $32.34_{-8.10}^{+4.03}$                   &  $33.83_{-8.99}^{+4.98}$                &    $24.14_{-1.99}^{+6.99}$        &   $30.98_{-7.21}^{+3.95}$   &     $28.58_{-5.58}^{+4.58}$     \\     
$ K $               &  mag                    &   $33.11_{-9.40}^{+7.63}$                   &  $35.33_{-11.14}^{+8.14}$                &    $23.66_{-1.92}^{+8.07}$        &   $31.09_{-7.82}^{+5.73}$   &     $28.29_{-5.73}^{+6.75}$     \\   
\hline \hline
       \end{tabular}
\label{tab-bay}
\end{center}
\end{table}

\begin{figure}[h]
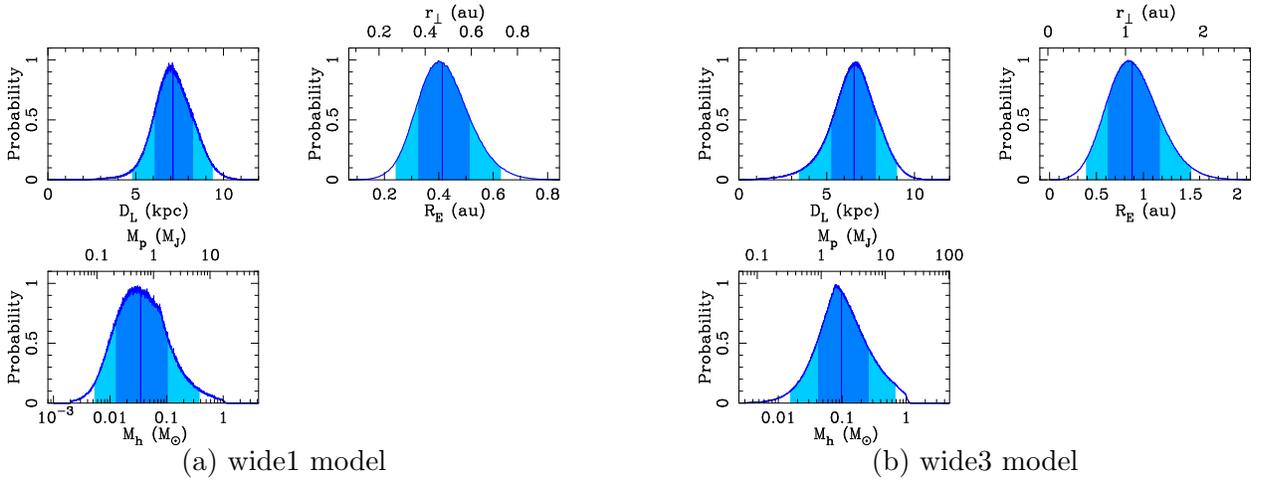

\begin{minipage}{0.5\hsize}
 	\centering
	\includegraphics[scale=0.30,angle=270]{kondo0419.1.ps}
	\hspace{1.6cm} (a)\ wide1\ model
\end{minipage}
\begin{minipage}{0.5\hsize}
 	\centering
	\includegraphics[scale=0.30,angle=270]{kondo0419.2.ps}
	\hspace{1.6cm} (b)\ wide3\ model
\end{minipage}
\caption{Probability distribution of lens properties for the wide1 model and the wide3 model by Bayesian analysis. The vertical blue lines show the median values. The dark-blue and the light-blue regions show the 68.3\% and 95.4\% confidence intervals.}
\label{fig-bay}
 \end{figure}

\begin{figure}[h]
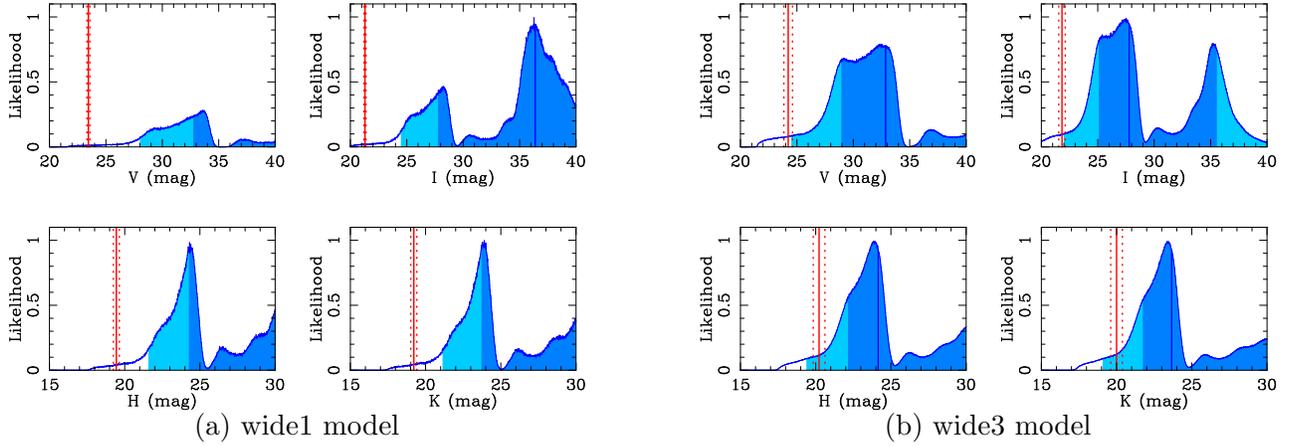

\begin{minipage}{0.5\hsize}
 	\centering
	\includegraphics[scale=0.30,angle=270]{kondo0419.4.ps}
	\hspace{1.6cm} (a)\ wide1 model
\end{minipage}
\begin{minipage}{0.5\hsize}
 	\centering
	\includegraphics[scale=0.30,angle=270]{kondo0419.5.ps}
	\hspace{1.6cm} (b)\ wide3 model
\end{minipage}
\caption{Probability distribution of $I$-, $V$-, $H$-, and $K$-band magnitudes with  extinction as a result of the Bayesian analysis. The dark-blue and the light-blue regions show the 68.3\% and 95.4\% confidence intervals, respectively. The vertical blue lines show the median values. The red vertical lines show the magnitudes of the source stars with extinction. As for the $I$-band and $V$-band, the source magnitude is derived from the light-curve modeling. As for the $H$-band and $K$-band, the source magnitude is derived from \citet{1995ApJS..101..117K} and \citet{2003A&A...402..701B}, and the extinction is estimated by using \citet{1989ApJ...345..245C}.}
\label{fig-bay2}
 \end{figure}

Table \ref{tab-bay} shows the summary of the lens physical parameters of each model, and 
we found that the results are divided into two types according to the mass of the host star.  The median value of the probability distribution of the host mass for all models except wide3 indicates the host star is a brown dwarf, and that of the wide3 indicates that the host star is a late M-dwarf.
As for the wide1 model, the lens system could be a gas giant with a mass of $M_{\rm p} = 0.60\ M_{\rm Jup}$ orbiting a brown dwarf with a mass of $M_{\rm h} = 0.03\ M_\odot$, located at $D_{\rm L} = 7.12\ {\rm kpc}$ from the Earth, and a projected separation from the host star of $r_\perp = 0.48\ {\rm au}$. If we assume a circular and randomly oriented orbit for the planet, the three-dimensional semi-major axis is expected to be $a_{\rm 3D} = 0.59\ {\rm au}$.
Assuming the wide3 model is correct, the lens system is likely a gas giant with a mass of $M_{\rm p} = 0.63\ M_{\rm Jup}$ orbiting an M-dwarf with a mass of $M_{\rm h} = 0.10\ M_\odot$, located at $D_{\rm L} = 6.57$ kpc from the Earth, and a projected separation of $r_\perp = 1.54\ {\rm au}$ and the three-dimensional semi-major axis is expected to be $a_{\rm 3D} = 1.92\ {\rm au}$.
Figure \ref{fig-bay} shows the probability distribution of the lens parameters of the wide1 model and the wide3 model. The dark and light-blue regions show the $1\sigma$ and $2\sigma$ confidence intervals, respectively, and the vertical blue lines show the median value. The probability distributions of $V$-, $I$-, and $H$-band magnitudes with extinction of the host star of the wide1 model and the wide3 model are also shown in Figure \ref{fig-bay2}.

\begin{figure}[h]
\begin{center}
   \includegraphics[scale=0.35, angle=270]{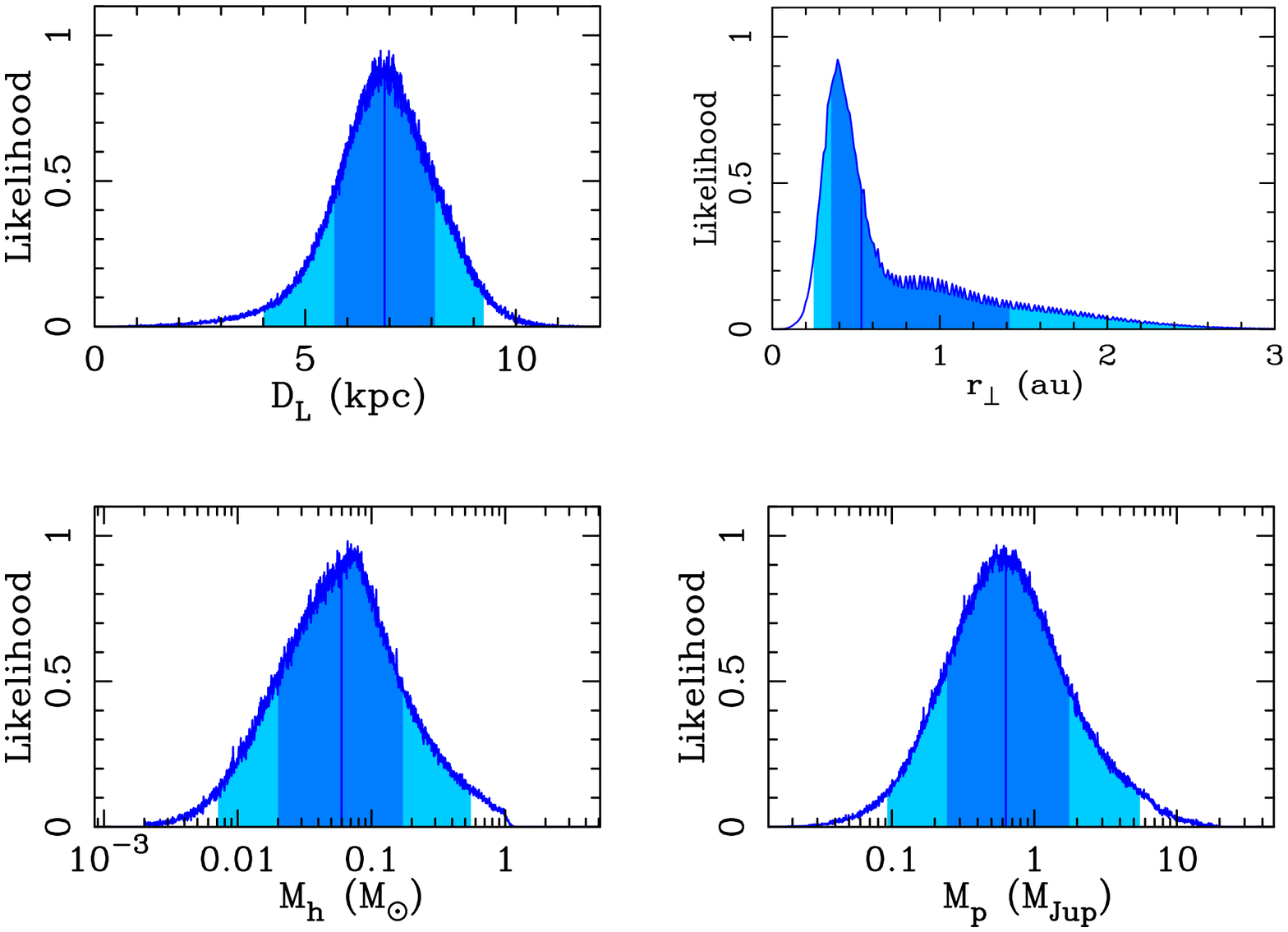}
   \caption{The result of combining the probability distributions of the lens properties of the five models. The vertical blue lines show the median value. The dark-blue and the light-blue regions show the 68.3\% and 95.4\% confidence intervals, respectively.}
\end{center}
\label{fig-bay3}
\end{figure}
Since the probability distributions of the lens physical parameters for all five models are consistent within $1\sigma$, we combined the distributions of these five models. Here, we combined these distributions with equal weight because all models are equally good within the uncertainty including the possible systematics. Figure \ref{fig-bay3} shows the combined probability distributions, which indicate that the lens system comprises a gas-giant planet with a mass of $M_{\rm p} = 0.63^{+1.13}_{-0.39}\ M_{\rm Jup}$ orbiting a brown dwarf with mass of $M_{\rm h} = 0.06^{+0.11}_{-0.04}\ M_\odot$ at $D_{\rm L} = 6.89^{+1.19}_{-1.19}\ {\rm kpc}$ and a projected separation of $r_\perp = 0.53^{+0.89}_{-0.18}\ {\rm au}$.
\\

\section{Discussion and Summary}
\label{sec-dis}
We analyzed the short duration microlensing event MOA-bin-29, which was found only after conducting an off-line analysis of the MOA database using data from $2006-2014$ (Sumi et al. 2019, in preparation). Although we found five competing solutions, all degenerate models have a planetary mass ratio $\sim10^{ -2 }$ and an Einstein radius crossing time of $~4-7$ days.
The angular Einstein radius estimated from the detection of the finite source effects was used to constrain the lens parameters for some models.
As a result of a Bayesian analysis, we found that the lens system is likely to be a gas giant orbiting a brown dwarf or a late M-dwarf.  
\\

Future high-resolution imaging with ground-based AO observations or space telescope could constrain the lens parameters \citep{2006ApJ...647L.171B, 2007ApJ...660..781B, 2015ApJ...808..169B, 2014ApJ...780...54B, 2015ApJ...808..170B, 2017AJ....154...59B, 2017AJ....154....3K,2018AJ....156..289B}.The source and the lens will be separated by $\sim 100$ mas for the wide1 model and $\sim 160$ mas for the wide3 model by 2025.
According to the Figure \ref{fig-bay2} and Table \ref{tab-bay}, the lens $K$-band magnitude with extinction would be $K \sim 33$ mag and $K \sim 24$ mag for the wide1 and the wide3 model, respectively. These are $\sim 13$ mag and $\sim 4$ mag fainter compared to the source for the wide1 and the wide3, respectively. This indicates that there is only small chance to detect the lens flux even if the lens and sources are separated by $\sim 100$ mas.
However the source flux could be detected using Keck AO \citep{2014ApJ...780...54B} or $JWST$ \citep{2006SSRv..123..485G} because the measured source magnitude with extinction is bright enough to be detected. 
The source flux for each model is not well determined by the light-curve modeling, so the improvement of the accuracy of the source flux would constrain the degenerate models.
\\

According to \citet{2016ApJ...833..145S}, the detection efficiency and the survey sensitivity of planetary systems with a short duration is relatively low, so only a few microlensing planets with a short duration have been found \citep{2014ApJ...785..155B}, and therefore 
any inferences drawn from these data have large statistical errors.
It is therefore important to increase the number of samples of these planets.
However the determination of lens properties for short duration events is difficult because in such a short event the measurement of a significant microlensing parallax effect is almost impossible. In order to solve the relation between  lens physical parameters, other constraints are important, for example follow-up observations with high-resolution imaging as described above. 
Measuring the space parallax is also a powerful way to determine the lens parameters for short duration events \citep{2017ApJ...840L...3S, 2019ApJ...871..179C}. Especially, Earth-L2 separation between the ground and the $WFIRST$ telescope \citep{2015arXiv150303757S} will be optimal for such short events with small Einstein radius.
\\

\begin{figure}[h]
\begin{center}
    \includegraphics[scale=0.8, angle=0]{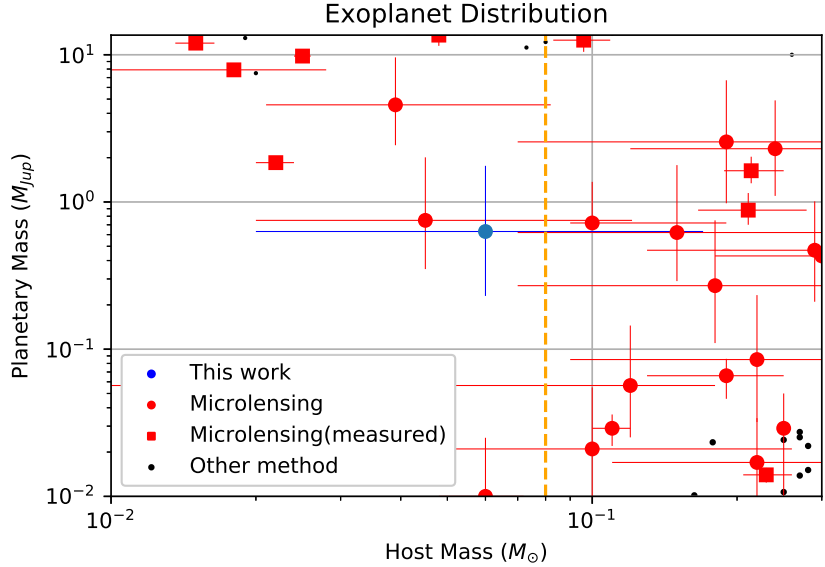}
   \caption{Mass distribution of a brown dwarf/a late M-dwarf hosting a gas giant. We choose systems with $0.01\ M_\odot < M_{\rm h} < 0.3\ M_\odot$,  $0.01\ M_{\rm Jup} < M_{\rm p} < 13.6\ M_{\rm Jup}$ from http://exoplanet.eu. The blue dot shows MOA-bin-29. The red dots show planets discovered by the microlensing method while black dots show those found by other methods. As for microlensing planets, square indicate that their masses are measured and circles indicate that their masses are estimated by a Bayesian analysis. The values of microlensing planet are from each paper. The orange line indicates a boundary of a brown dwarf/a late M-dwarf host star ($M_{\rm h} = 0.08\ M_\odot$).}
\end{center}
\label{fig-exoplanet}
\end{figure}

The formation theory of a gas giant orbiting a brown dwarf and even that for a brown dwarf itself is still ambiguous.
As a result of analyzing the event MOA-bin-29, we found the lens system is most likely to be a gas giant orbiting a brown dwarf, which is hard to form according to the core accretion theory.
Therefore more and more accurate mass measurements of such systems are required.
The microlensing method is a powerful way to detect these systems and
some of such events have been discovered by the microlensing method \citep{2013ApJ...778...38H, 2018ApJ...858..107A, 2018AJ....155..219J, 2018AJ....156..208J}.
Figure \ref{fig-exoplanet} shows the mass distribution of discovered systems with a brown dwarf/a late M-dwarf hosting a gas giant ($0.01\ M_\odot\ <\ M_{\rm h}\ <\ 0.3\ M_\odot$,  $0.01\ M_{\rm Jup}\ <\ M_{\rm p}\ <\ 13.6\ M_{\rm Jup}$).
For most of these events, the lens physical parameters such as mass are derived through a Bayesian analysis, which uses the mass function as a prior. The mass probability distribution of low-mass host stars ($M_{\rm h}\ <\ 0.1\ M_\odot$) depends strongly on the shape of the mass function which has large uncertainty.
In \citet{2018ApJ...858..107A}, the lens parameters are directly derived by combining measurements from Earth and from the $Spitzer$ telescope \citep{2014sptz.prop11006G}. The increase in number of the mass measurements by the space parallax effect and high-resolution images are greatly anticipated by $WFIRST$ satellite in the near future, and would contribute significantly to clarification of the formation theory of a gas giant around a low-mass star.
\\

T.S. acknowledges the financial support from the JSPS, JSPS23103002, JSPS24253004, and JSPS26247023. The MOA project is supported by the grant JSPS25103508 and 23340064. D.P.B., A.B., and D.S. were supported by NASA through grant NASA-NNX12AF54G. Work by N.K. is supported by JSPS KAKENHI grant No. JP15J01676. Work by A.F. is supported by JSPS KAKENHI grant No. JP17H02871. N.J.R. is a Royal Society of New Zealand Rutherford Discovery Fellow. A.S. is a University of Auckland Doctoral Scholar. The OGLE project has received funding from the National Science Centre, Poland, grant MAESTRO 2014/14/A/ST9/00121 to AU.

\appendix
\label{app-ba}
\section{Systematic Photometry Errors on the Baseline}
In Section 3.3, our grid search found the five competing models. The $\chi^2$ differences between these different 
models are primarily due to the main light-curve peak, but several small bumps in the light-curve baseline 
also contribute.
The bump at $\rm HJD'$ $\sim$ $3929.7-3930.0$ is most likely to be a real caustic feature. The
caustic geometries of the wide1 and wide2 models require a feature at approximately this time, and the significance
of this feature is $\Delta\chi^2 = 15.1$. The other small bumps (Figure \ref{fig-light2} (c) and (d)) could be due to 
systematic photometry errors. These are further from the peak, and the caustic geometry of the event does not
require that the source trajectory will encounter them, particularly if realistic microlensing parallax and orbital 
motion are included. These features 
have a lower significance of $\Delta\chi^2 = 14.0$ and $\Delta\chi^2 = 7.3$ for features (c) and (d), respectively.
On the other hand, these data are not associated with 
bad seeing, high airmass, or large PSF fit $\chi^2$ in the fit to the difference images. 
Visual inspection of the images does not show any obvious problems. 
We now investigate if the existence of these two bumps might 
affect the final results as follows. While we are not certain if these bumps are systematic errors or real features, 
we can estimate the possible uncertainty by this exercise.
\\

In the best-fit models, we removed data points during night that contain the small bump causing the $\chi^2$ differences. Here, we removed all of the data points in the night because we cannot set a clear boarder line to cut these specific data points in question because there are a lot of similar $2-3\sigma$ outliers. Then, we conducted a grid search for the best-fit models. We repeated this process three times until there are no $\chi^2$ differences from the small bumps in the baseline. After this iteration, we found only two new local minima in addition to the models, which are almost same as the original best-fit five models. These new models are within the range of parameters of the original five models, $6.0\times10^{ -3 } \leq q \leq 2.0\times10^{ -2 }$ and $0.5 \leq s \leq 1.8$. The distributions of the MCMC chains of these new local minima also roughly overlap to those of original models.
Thus, the bumps that we decided are likely to be due to systematics influence only the best fit models. The best fit models are decided by very small changes in $\chi^2$. However, the inclusion, or not, of these bumps due to systematics is irrelevant because they do not affect the MCMC distributions, which show the real determination of the features of the light curve.
\\

\begin{figure}[h]
\begin{center}
   \includegraphics[scale=0.35, angle=270]{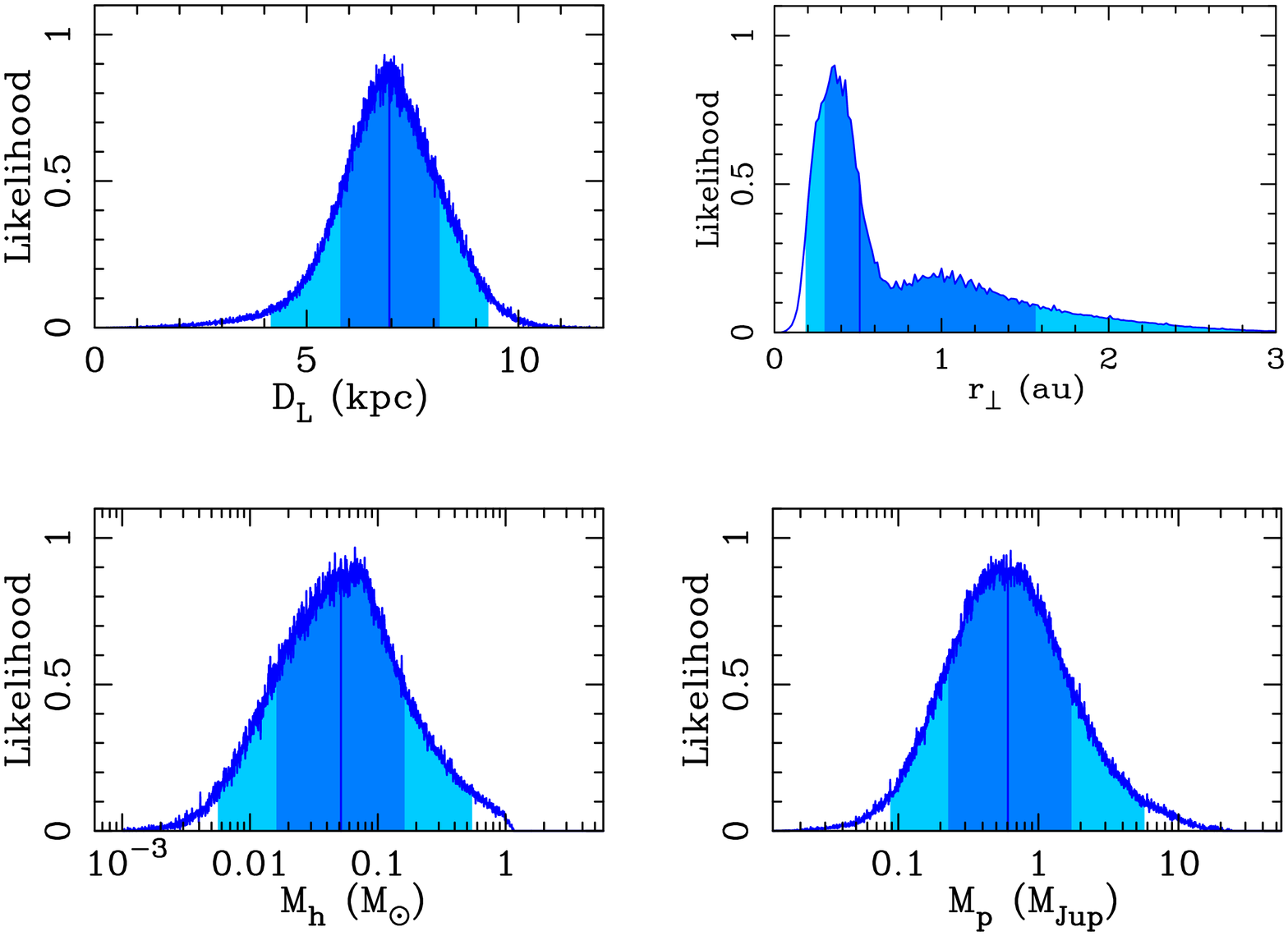}
   \caption{The result of combining the probability distribution of the lens properties of five models and three local minima. The vertical blue lines show the median value. The dark-blue and the light-blue regions show the 68.3\% and 95.4\% confidence intervals, respectively.}
\end{center}
\label{app-bay-fig}
\end{figure}

Furthermore, in order to check if the final probability distribution of the lens properties would change with these new local minima, we conducted a Bayesian analysis including these new models.
Figure \ref{app-bay-fig} shows the combined probability distributions which include the probability distribution of three new local minima and indicates the lens system comprise a gas-giant planet with a mass of $M_{\rm p} = 0.61^{+1.12}_{-0.38}\ M_{\rm Jup}$ orbiting a brown dwarf with mass of $M_{\rm h} = 0.05^{+0.11}_{-0.04}\ M_\odot$ at $D_{\rm L} = 6.95^{+1.19}_{-1.16}\ {\rm kpc}$ and a projected separation of $r_\perp = 0.51^{+1.05}_{-0.21}\ {\rm au}$.
Thhese probability distributions of lens physical parameters are consistent with the original result in Section \ref{sec-bay}.
We get a similar result when we conduct the same analysis, assuming the bump around $\rm HJD'$ $\sim$ $3929.7-3930.0$ is due to the systematics, to be safe.
\\

Therefore, we use these original five models without removing any data points but recognize the range of these models as uncertain. Thus, we treat these five models equally, not weighting by $\Delta \chi^2$, in a Bayesian analysis, in Section \ref{sec-bay}.

{}


\end{document}